\documentclass[ twocolumn, trackchanges]{aastex701}
\nolinenumbers

\usepackage{gensymb}
\usepackage{color}
\definecolor{maroon}{RGB}{139,0,0}
\usepackage{comment}

\newcounter{qnumber}

\begin{document}

\title{TOI-2665 b: A Highly Irradiated Giant in the Hot Neptune Desert\footnote{This paper includes data gathered with the 6.5 meter Magellan Telescopes located at Las Campanas Observatory, Chile.}}

\author[orcid=0009-0001-8823-8843]{Darya Daniels} 
\affiliation{Department of Physics and Astronomy, University of British Columbia, 6224 Agricultural Road, Vancouver, BC V6T 1Z1, Canada}
\email[show]{ddanie04@student.ubc.ca}

\author[orcid=0000-0001-9269-8060]{Michelle Kunimoto}
\affiliation{Department of Physics and Astronomy, University of British Columbia, 6224 Agricultural Road, Vancouver, BC V6T 1Z1, Canada}
\email{mkuni@phas.ubc.ca}

\author[orcid=0000-0003-2733-8725]{James S. Jenkins}
\affiliation{Instituto de Estudios Astrof\'isicos, Universidad Diego Portales, Av. Ej\'ercito 441, Santiago, Chile}
\affiliation{Centro de Astrof\'isica y Tecnolog\'ias Afines (CATA), Casilla 36-D, Santiago, Chile}
\email{james.jenkins@mail.udp.cl}

\author[orcid=0000-0002-5619-2502]{Douglas Alves}
\affiliation{Departamento de Astronomía, Universidad de Chile, Casilla 36-D, Santiago, Chile}
\email{douglasalvesastro12@gmail.com}

\author[orcid=0000-0002-3481-9052]{Keivan G. Stassun}
\affiliation{Department of Physics and Astronomy, Vanderbilt University, 6301 Stevenson Center Lane, Nashville, TN 37235, USA}
\affiliation{Department of Physics, Fisk University, 1000 17th Avenue North, Nashville, TN 37208, USA}
\email{keivan.stassun@Vanderbilt.Edu}

\author[orcid=0000-0001-7961-3907]{Samuel~W.~Yee}
\altaffiliation{51 Pegasi b Fellow}
\affiliation{Department of Physics \& Astronomy, University of California Los Angeles, Los Angeles, CA 90095, USA}
\email{syee@astro.ucla.edu}

\author[orcid=0000-0003-1305-3761]{R. Paul Butler}
\affil{Earth and Planets Laboratory, Carnegie Institution for Science, 5241 Broad Branch Rd NW, Washington, DC 20015, USA  }
\email{bluaper@gmail.com}

\author[orcid=0000-0002-5226-787X]{Jeffrey D. Crane}
\affiliation{The Observatories of the Carnegie Institution for Science, 813 Santa Barbara Street, Pasadena, CA, 91101, USA}
\email{crane@carnegiescience.edu}

\author[0000-0001-6039-0555]{Fabo Feng}
\affiliation{ State Key Laboratory of Dark Matter Physics, Tsung-Dao Lee Institute}
\affil{School of Physics and Astronomy, Shanghai Jiao Tong University, Shanghai 201210, People’s Republic of China}
\email{ffeng@sjtu.edu.cn}

\author[orcid=0000-0002-2532-2853]{Steve B. Howell}
\affiliation{NASA Ames Research Center, Moffett Field, CA 94035, USA}
\email{steve.b.howell@nasa.gov}

\author[0000-0001-8401-4300]{Shubham Kanodia}
\affil{Earth and Planets Laboratory, Carnegie Institution for Science, 5241 Broad Branch Rd NW, Washington, DC 20015, USA  }
\email{skanodia@carnegiescience.edu}

\author[orcid=0000-0003-0412-9664]{David Osip}
\affiliation{
Carnegie Science
Las Campanas Observatory }
\email{dosip@carnegiescience.edu}

\author[orcid=0000-0002-3940-2360]{Don Radford}
\affiliation{Brierfield Observatory, Bowral, NSW, Australia}
\email{dradford63@gmail.com}

\author[orcid=0000-0002-8681-6136]{ Stephen A. Shectman}
\affiliation{The Observatories of the Carnegie Institution for Science, 813 Santa Barbara Street, Pasadena, CA, 91101, USA}
\email{sshec2@gmail.com}

\author[orcid=0009-0008-2801-5040]{Johanna K. Teske}
\affiliation{Earth and Planets Laboratory, Carnegie Institution for Science, 5241 Broad Branch Rd NW, Washington, DC 20015, USA  }
\affil{The Observatories of the Carnegie Institution for Science, 813 Santa Barbara Street, Pasadena, CA, 91101, USA}
\email{jteske@carnegiescience.edu}

\author[0000-0002-0619-7639]{Carl Ziegler}
\affil{Department of Physics, Engineering and Astronomy, Stephen F. Austin State University, 1936 North St, Nacogdoches, TX 75962, USA}
\email{carlziegler@gmail.com}

\begin{abstract}
We report the confirmation of TOI-2665 b, a close-in sub-Saturn ($P=2.85$ days, $R_p=7.00_{-0.15}^{+0.16}~R_{\oplus}$) just inside the hot Neptune desert around a metal-rich G dwarf, observed by the Transiting Exoplanet Survey Satellite (TESS). We make use of the Carnegie Planet Finder Spectrograph for precise radial velocity measurements to measure its mass with $13.5\sigma$ precision ($M_p= 57.0_{-4.2}^{+4.2} ~M_{\oplus}$). The low density of TOI-2665 b ($\rho_{p} \sim 0.9~\mathrm{gcm}^{-3}$) implies the existence of a significant gaseous envelope that has survived in the hot Neptune desert despite high irradiation ($S=1068~S_{\oplus}$), making it an intriguing target for understanding the formation and evolution of desert-dwelling giants. The relatively bright host star ($V=11.0$ mag,  $J=9.8$ mag) and extensive atmosphere make TOI-2665 b a promising target for atmospheric characterization.
\end{abstract}


\section{Introduction} 

The hot Neptune desert is a striking feature of the exoplanet population, defined by a dearth of sub-Neptune to sub-Jovian-sized planets ($R_{p} \sim [2, 10]~R_{\oplus}$) in close-in orbits ($P \lesssim 3$ days) \citep{SzaboKiss2011, Mass-period2011}. The lack of hot Neptunes creates a wedge like gap surrounded by super-Earths and hot Jupiters in both period-radius and period-mass space. The lower edge of the desert is broadly accepted to be sculpted by photoevaporation \cite[e.g., ][]{KurokawaAndNakamoto2014, OwenLai2018}, by which planets lose their atmospheres due to high incident radiation, leaving behind bare cores. The upper edge is likely sculpted by the tidal disruption of gas giants following high eccentricity migration \cite[e.g., ][]{OwenLai2018, Zanazzi2026}.

Recent discoveries especially by the Transiting Exoplanet Survey Satellite (TESS) mission \citep{Ricker2015TESS}, however, have shown that the desert is not completely empty. While some hot Neptunes can be explained by being very young or recently migrated, and are therefore currently experiencing high levels of atmospheric mass-loss \cite[e.g. V1298 Tau b;][]{V1298Taub}, others have more persistently survived in the desert over time. The existence of these more persistent desert-dwellers challenges conventional theories of planet formation and evolution.

\citet{VissapragadaBehmard2025} outlined three possible origin pathways for long-standing desert-dwellers. First, some hot Neptunes may have formed similar to other Neptunes, but are simply unusually resistant to photoevaporation such as by having extremely metal-rich atmospheres \cite[e.g., ][]{OwenMurray-Clay2018.480.2206O, RadicaCoulombe2024}. Others may have formed ``bottom up'' via planet-planet collisions, producing planets with extremely high core densities and masses but small envelopes resilient to photoevaporation \cite[e.g., TOI-1853 b;][]{NaponielloTOI1853}. Finally, some desert-dwellers may have formed ``top down'' as hot Jupiters which have since lost significant portions of their envelopes. \citet{VissapragadaBehmard2025} concluded that the ``top down'' pathway may be the most dominant origin due to observing typically high metallicities among hot Neptune stellar hosts, given that high metallicity is known to correlate with hot Jupiter abundance \cite[e.g.][]{Fischer2005,Petigura2018}. If most hot Neptunes are indeed ``hot Jupiters gone wrong,'' then atmospheric characterization of such objects could offer insights into the interiors of giant planets. 

In this work we report the discovery and confirmation of TOI-2665 b, a hot sub-Saturn falling within the hot Neptune desert. The combination of its low density ($0.9$ gcm $^{-3}$) and metal-rich host star ([Fe/H] = $0.23^{+0.05}_{-0.06}$) are consistent with the interpretation that the planet is an ex-hot Jupiter, and its extensive atmosphere make it an attractive target for characterization of highly irradiated hot Neptunes. In \S\ref{sec:observations}, we discuss the photometry, high-resolution imaging, and spectroscopic observations that allowed for the discovery and confirmation of the system. In \S\ref{sec:star} and \S\ref{sec:planet} we discuss determination of the stellar and planetary parameters respectively. Finally in \S\ref{sec:discussion} we discuss our findings and implications of our results in the context of the hot Neptune desert.

\section{Observations}\label{sec:observations}

\subsection{TESS Discovery Photometry}
TESS is a NASA- and MIT-led space telescope that monitors the whole sky searching for transiting exoplanets, especially around the brightest stars. TESS uses four identical cameras with a $24{\degree} \times 24{\degree}$ field of view that together survey a $24{\degree} \times 96{\degree}$ strip of sky. The full images obtained by CCDs, known as Full-Frame Images (FFIs), are currently obtained at 200s cadence. Previously, FFIs were obtained at 1800s cadence (Sectors 1 -- 26) and 600s cadence (Sectors 27 -- 54).

TOI-2665 (TIC-186936449) was observed by TESS in Sectors 8, 35, 62, and 89. A 2.85-day planet candidate was first detected using Sector 8 FFI data by the neural network-based planet search described by \citet{Olmschenk2021}, and was submitted as a Community TESS Object of Interest (CTOI) on 2021 Febuary 08. MIT's Quick Look Pipeline \cite[QLP;][]{Huang2020,Kunimoto2021} later uncovered the same signal with FFI light curves from Sectors 8 and 35, leading to the alert of the candidate as TOI-2665.01 on 2021 May 06.

\begin{figure*}
    \centering
    \includegraphics[width=\linewidth]{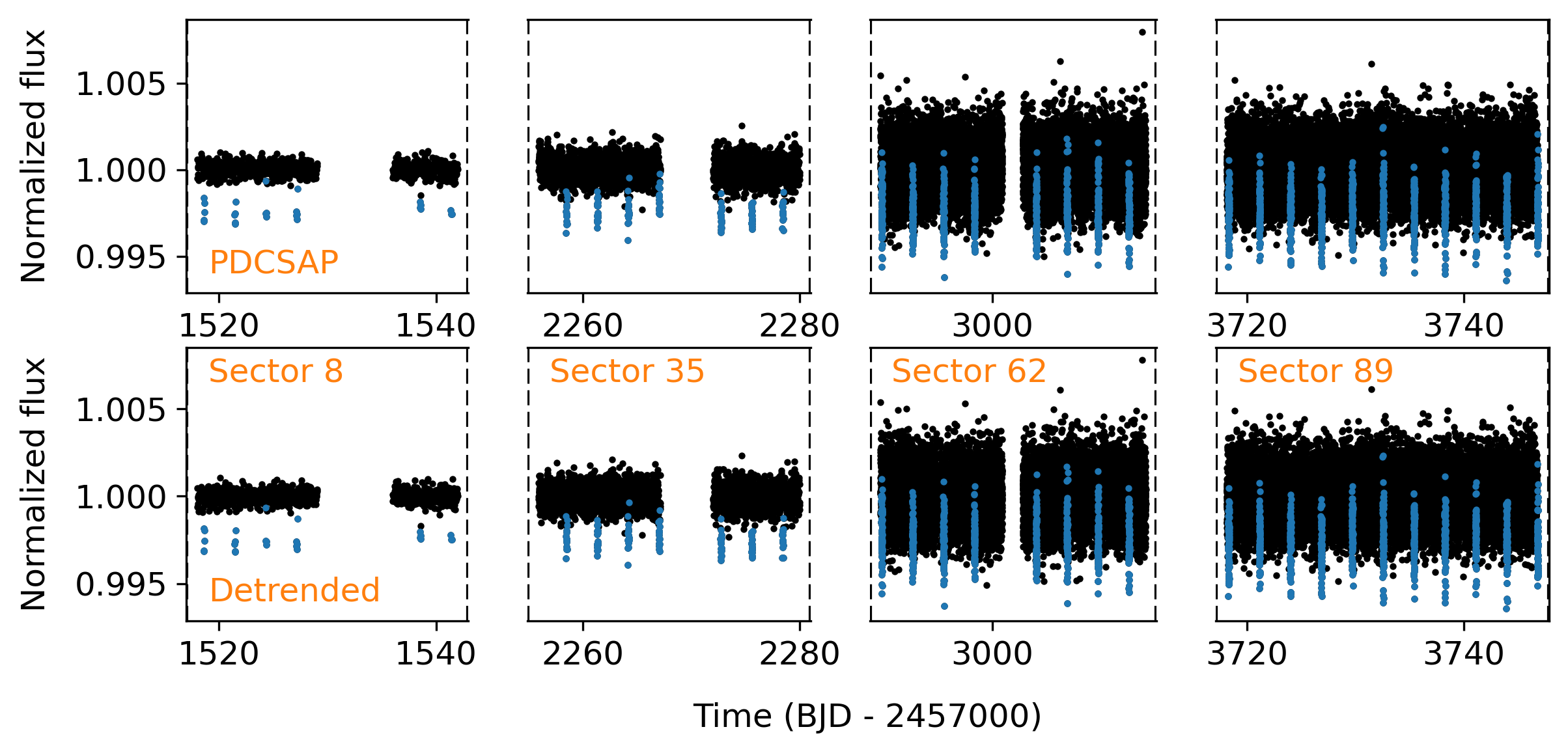}
    \caption{SPOC PDCSAP light curve for TOI-2665 from Sectors 8, 35, 62, and 89 before detrending (top) and after detrending (bottom). The transits of TOI-2665.01 are indicated by blue dots.}\label{fig:lightcurves}
\end{figure*}

For this work, we downloaded light curves produced by the TESS Science Processing Operations Center \cite[SPOC;][]{SPOC} from the Mikulski Archive for Space Telescopes (MAST).\footnote{https://mast.stsci.edu} Specifically, we obtained FFI light curves from Sectors 8 and 35 \citep{Caldwell2020}, and 2-minute cadence light curves from Sectors 62 and 89. We used the Presearch Data Conditioning Simple Aperture Photometry \cite[PDCSAP;][]{Smith2012,Stumpe2012,Stumpe2014} light curves with all data points with a nonzero quality flag removed. As shown in Figure \ref{fig:lightcurves}, the PDCSAP light curve is relatively flat outside of the transits, with no detectable rotation period visible on the observations. Nevertheless, we further removed low-frequency trends using the biweight time-windowed slider implemented in the \texttt{wotan} Python package \citep{Wotan} with a window of 0.5 days. We ran a blind search for periods between 1 and 10 days on the combined multi-sector light curve using the Box Least-Squares \cite[BLS;][]{Kovacs2002} algorithm implemented in \texttt{cuvarbase} \citep{Hoffman2022} and recovered one planet candidate with properties consistent with the TOI: an orbital period of $P = 2.85$ days, a transit epoch of $T_{0} = 1518.631$ days (BJD$\mathrm{_{TDB}}$ - 2457000), and a duration of 2.4 hours. We did not uncover any new candidates after masking out this candidate during further searches of the data.

Because detrending can distort transit shapes, we re-detrended the light curve after recovering the TOI such that the algorithm ignores cadences within one transit duration of the transit mid-points when determining the parameters of the trend functions. We use this re-detrended light curve (see Figure \ref{fig:lightcurves}) for the remainder of this paper.

\subsection{Ground-Based Photometry}

\begin{figure*}
    \centering
    \includegraphics[width=0.8\linewidth]{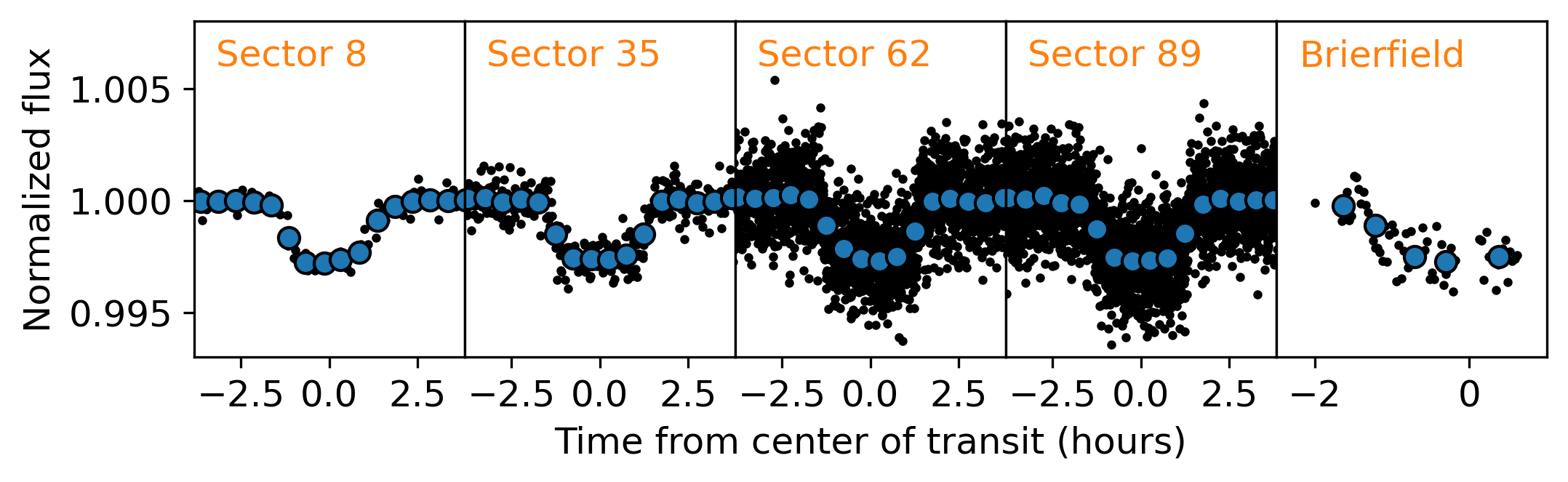}
    \caption{Phase-folded transits of TOI-2665.01 from each TESS sector as well as one ingress recovered from the ground at Brierfield Observatory. All transits show consistent depths. Blue datapoints show the light curve binned every 30 minutes.}\label{fig:all_transits}
\end{figure*}

We observed an ingress of the transit in the \textit{Rc}-band on 2025 February 17 from Brierfield Observatory near Bowral, New South Wales, Australia. The 0.36 m telescope is equipped with a $4096\times4096$ Moravian 16803 camera. At a pixel scale of $0.735^{\prime\prime}~\mathrm{px}^{-1}$, the photometric data were extracted using \texttt{AstroImageJ} \citep{AstroImageJ} with a circular 9 pixel ($6.6^{\prime\prime}$) photometric aperture. The transit has a depth consistent with the TESS transits ($\sim2.5$ ppt, Figure \ref{fig:all_transits}). This time series is available on the Exoplanet Follow-up Observing Program website\footnote{\url{https://exofop.ipac.caltech.edu}} \cite[ExoFOP;][]{ExoFOP}.

\subsection{High-Resolution Imaging}

We obtained high-resolution images of TOI-2665 to search for nearby stars (either bound or chance-aligned) which could be false positive sources for the candidates. These companions can also dilute the TESS photometry, resulting in under-estimated planet radii.

\begin{figure}
    \centering
    \includegraphics[width=\linewidth]{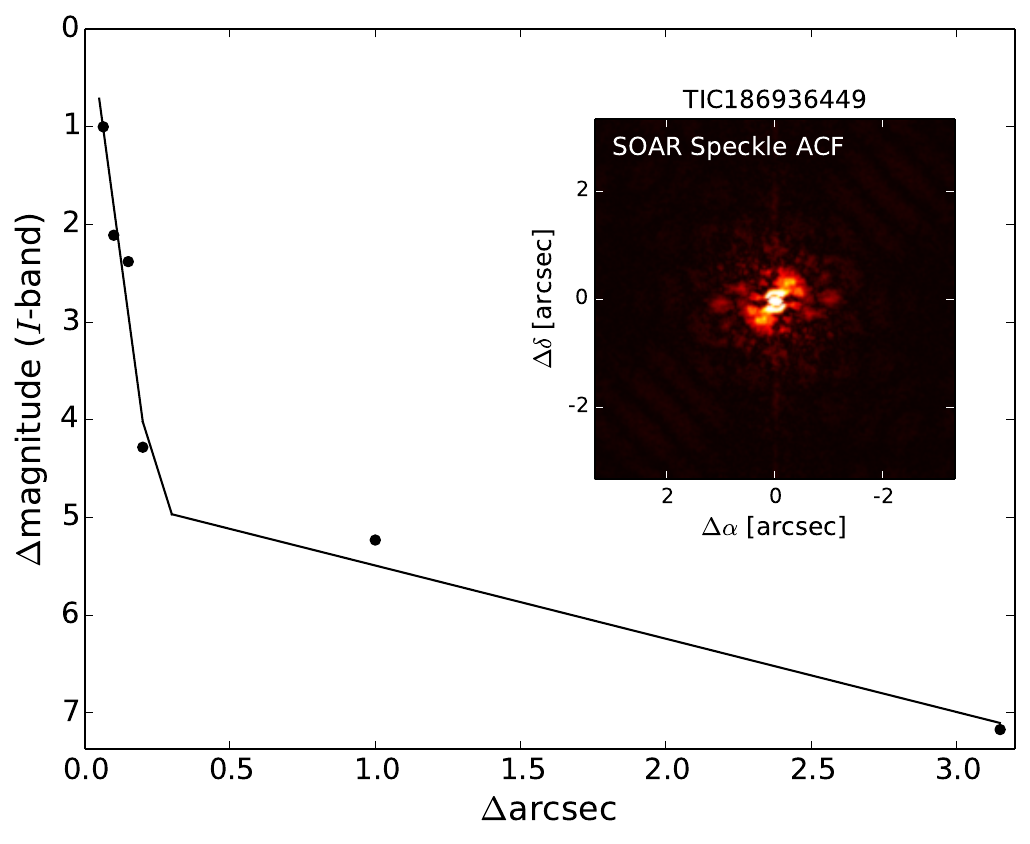}
    \includegraphics[width=\linewidth]{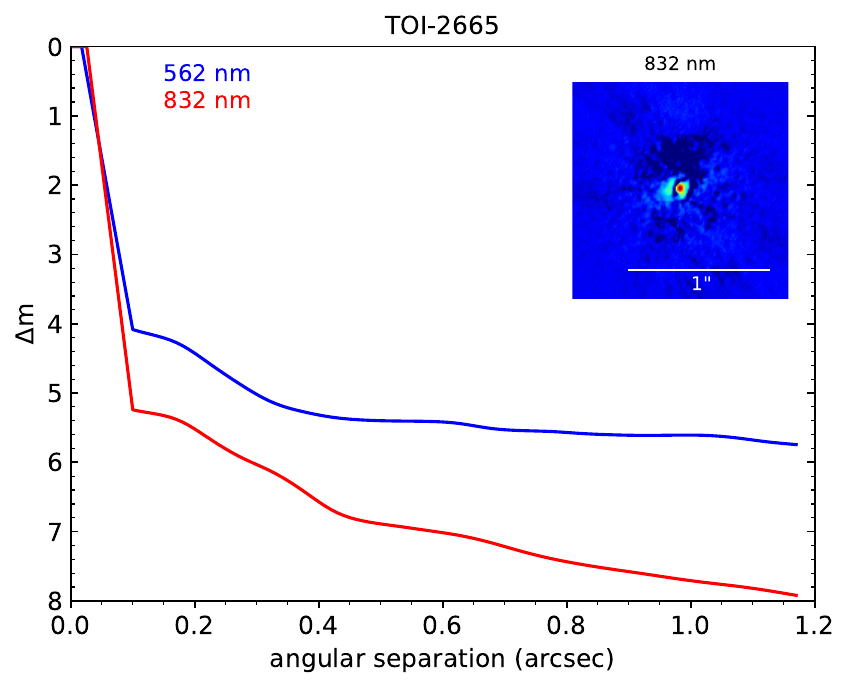}
    \caption{Contrast curves and high-resolution images from SOAR/HRCam speckle observations (top) and Gemini-S/Zorro speckle observations (bottom). No stellar companions were detected in any observations.}
    \label{fig:HRI}
\end{figure}

TOI-2665 was observed with speckle imaging using HRCam on the 4.1 m Southern Astrophysical Research (SOAR) telescope \citep{HRCam} on 2021 November 20. The images were collected at 879 nm with a pixel scale of $0.01575^{\prime\prime}~\mathrm{px}^{-1}$. TOI-2665 was also observed with speckle imaging by the Zorro instrument on the 8 m Gemini-South telescope \citep{Scott2021} on 2025 March 11. The images were collected at 562 and 832 nm with a pixel scale of $0.01^{\prime\prime}~\mathrm{px}^{-1}$. The data were reduced using the standard Fourier techniques as outlined in \citet{Howell2011}. To within the angular and magnitude limits achieved by the observations, no stellar companions were detected within the sensitivity limits of SOAR/HRCam ($\Delta m = 7.2$ at $1^{\prime\prime}$) or Gemini-S/Zorro ($\Delta m = 5.4$ and $\Delta m = 6.89$ at $0.5^{\prime\prime}$ for 562 and 832nm, respectively), as shown in Figure \ref{fig:HRI}. For the Gemini observations, the angular limits achieved correspond to spatial distances of 4.4 to 266 au.

\subsection{High-Resolution Spectroscopy}\label{sec:rvs}

We obtained a reconnaissance spectrum of TOI-2665 on 2021 December 30 using the MINERVA-Australis array. MINERVA-Australis is a facility consisting of multiple 0.7m telescopes feeding a high-resolution optical spectrograph ($R \sim 80,000$, 480 - 630 nm), located at Mount Kent Observatory in Queensland, Australia \citep{Addison2019}. The spectrum suggested that TOI-2665 is a moderately rotating star suitable for high-precision radial velocity measurements, with a projected rotational velocity of $v \sin{i} \sim 6$ km/s.

\begin{figure}
    \centering
    \includegraphics[width=0.9\linewidth]{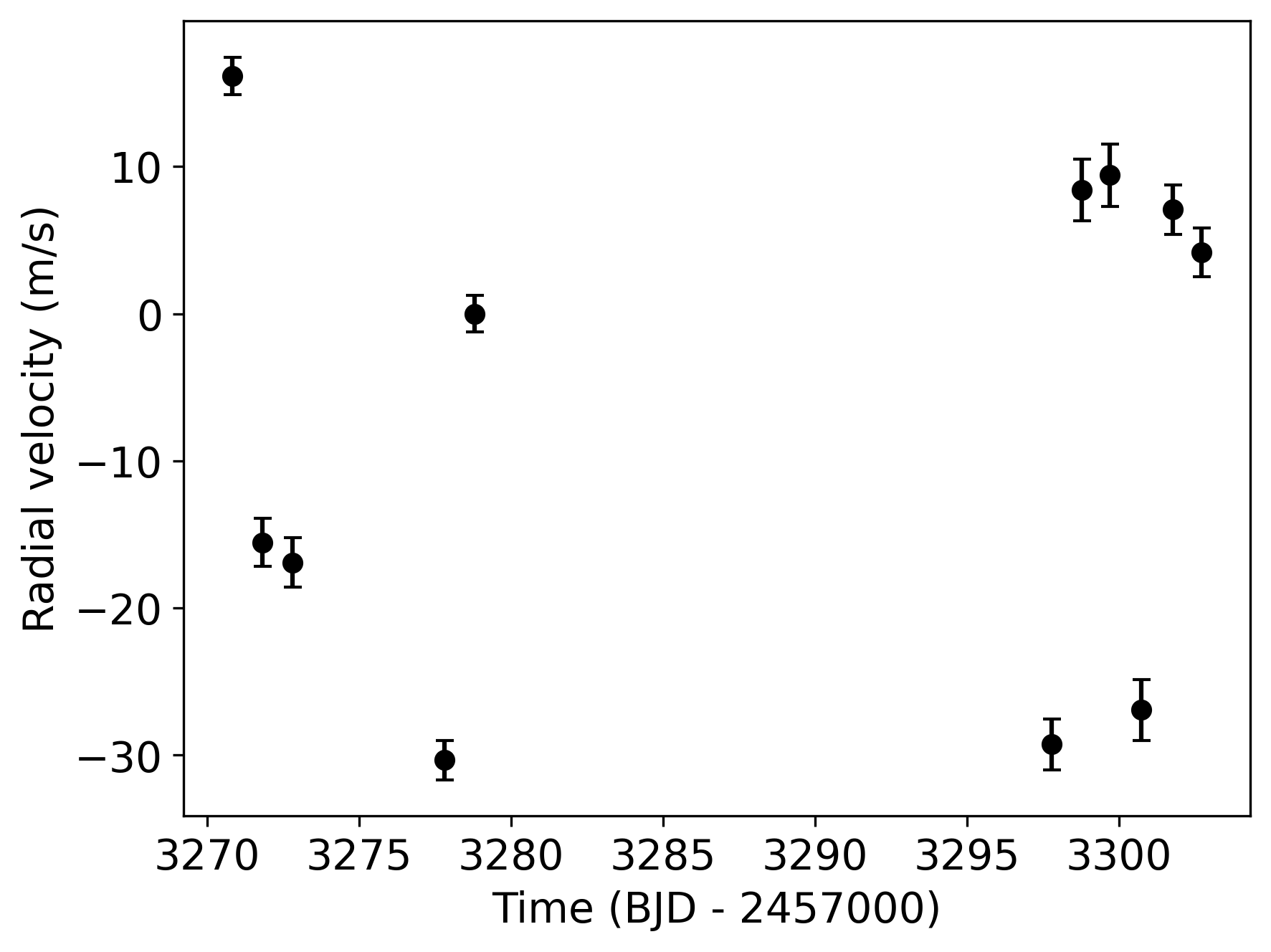}
    \caption{RV measurements from Magellan II/PFS. The RVs do not feature large variations corresponding to stellar-mass companions, and they are sufficiently precise to measure the mass of TOI-2665.01 and confirm its planetary nature (\S\ref{sec:planet}).}\label{fig:pfs}
\end{figure}

We later observed TOI-2665 with the Carnegie Planet Finder Spectrograph \cite[PFS;][]{Crane2006, Crane2008, Crane2010} on the 6.5m Magellan II Telescope at Las Campanas Observatory, between 2023 November 22 and 2023 December 24. PFS is a high-resolution optical echelle spectrograph with a total wavelength coverage of $391 - 734$ nm. The use of an iodine cell to measure RVs results in a wavelength range used for RV derivation of $500 - 620$ nm. RVs were extracted following the technique described by \citet{Butler1996}. We obtained 11 RVs with a mean uncertainty of $1.67~\mathrm{ms}^{-1}$ (Figure \ref{fig:pfs}), sufficiently precise for mass determination and subsequent confirmation of TOI-2665.01 as a planet (\S\ref{sec:planet}). The RVs range from $-30$ to $17~\mathrm{ms}^{-1}$ and thus lack large-scale variations, indicating no detectable stellar-mass companions. The full list of derived RVs are provided in Table \ref{tab:rvs}.

\begin{table}[]
    \centering
    \caption{Radial velocity measurements from Magellan II/PFS.}
    \begin{tabular}{ccc}
    \hline\hline
        Time & RV & RV error \\
        (\textbf{BJD$\mathrm{_{TDB}}$} - 2457000) & (ms$^{-1}$) & (ms$^{-1}$) \\
        \hline
        3270.82675 & 16.16 & 1.27 \\
        3271.82336 & -15.55 & 1.63 \\
        3272.82857 & -16.9 & 1.68 \\
        3277.80628 & -30.33 & 1.34 \\
        3278.80854 & 0.0 & 1.25 \\
        3297.78237 & -29.26 & 1.72 \\
        3298.79133 & 8.41 & 2.09 \\
        3299.69885 & 9.41 & 2.1 \\
        3300.72831 & -26.92 & 2.08 \\
        3301.76690 & 7.08 & 1.67 \\
        3302.72657 & 4.17 & 1.64 \\
    \end{tabular}
    \label{tab:rvs}
\end{table}

\section{Stellar Parameters}\label{sec:star}
The TICv8.2 gives stellar properties for TIC-186936449 of $M_{\star}=1.15 \pm 0.16~M_\odot$, $R_{\star}=1.302\pm 0.060~R_\odot$, $\log g = 4.269 \pm 0.082~(\mathrm{cgs})$, and $T_{\mathrm{eff}}=6112\pm 132$ K, indicating that TOI-2665 is an early G dwarf. We use several independent methods to measure and derive a range of stellar parameters for TOI-2665, and provide a summary in Table \ref{tab:StellarParams}.

\begin{table*}[]
\centering
\caption{Literature and Derived Stellar Parameters for TIC-186936449 (TOI-2665)}\label{tab:StellarParams}
\begin{tabular}{l|l|l|l}
\hline\hline
Parameter & Value & Description & Source \\
\hline
\multicolumn{4}{c}{\emph{Astrometric Parameters}} \\
\hline
$\alpha$ & 08:56:38.149 & Right ascension (J2016) & Gaia DR3 \\
$\delta$ & -34:37:42.51 & Declination (J2016)& Gaia DR3 \\
$\bar{\omega}$ & $4.4837\pm 0.0117
$ & Parallax (mas) &  Gaia DR3 \\
$\mu_\alpha$ & $ -25.6531 \pm 0.0093$ & Proper motion R.A (mas yr$^{-1}$) & Gaia DR3 \\
$\mu_\delta$ & $ 0.983 \pm 0.012$  & Proper motion decl. (mas yr$^{-1}$) & Gaia DR3 \\
RUWE & 0.945 & Renormalized Unit Weight Error & Gaia DR3 \\
\hline
\multicolumn{4}{c}{\emph{Photometric Parameters}} \\
\hline
$T$ & $10.3716 \pm 0.006$ & TESS band magnitude (mag) & TICv8.2 \\
$B$ & $11.584 \pm 0.102$ & $B$ band magnitude (mag) & UCAC4 \\
$V$ & $11.016 \pm 0.007$ & $V$ band magnitude (mag) & UCAC4 \\
$G$ & $10.80817 \pm 0.000093$ & Gaia $G$ band magnitude (mag) & Gaia DR3 \\
$J$ & $9.8 \pm 0.026$ & $J$ band magnitude (mag) & 2MASS \\
$H$ & $9.508 \pm 0.022$ & $H$ band magnitude (mag) & 2MASS \\
$K$ & $9.452 \pm 0.021$ & $K$ band magnitude (mag) & 2MASS \\
$W1$ & $9.391 \pm 0.023$ & WISE $3.4\mu$m band magnitude (mag) & WISE \\
$W2$ & $9.448 \pm 0.02$ & WISE $4.6\mu$m band magnitude (mag) & WISE \\
$W3$ & $9.394 \pm 0.032$ & WISE $12\mu$m band magnitude (mag) & WISE \\
\hline
\multicolumn{4}{c}{\emph{Spectroscopic Parameters}} \\
\hline
$M_\star$ &  $1.12^{+0.07}_{-0.05}$  & Stellar mass ($M_\odot$) & PFS spectral analysis (\S\ref{sec:spec}) \\
$T_{\mathrm{eff}}$ & $5864 \pm 50$ & Effective temperature (K) & PFS spectral analysis \\
$\log g$ & $4.17 \pm 0.09$ & Surface gravity (cgs) & PFS spectral analysis \\

[Fe/H] & $0.23\pm 0.05$ & Metallicity (dex) & PFS spectral analysis \\
$v\sin i$ & $5.00\pm0.28$ & Projected rotational velocity (km s$^{-1}$)& PFS spectral analysis \\
$\mathbf{\gamma}$ & $58.38\pm0.24$ & Barycentric radial velocity (km s$^{-1}$) & Gaia DR3 \\

\hline
\multicolumn{4}{c}{\emph{Final Derived Parameters}} \\
\hline
$R_\star$ & $1.302_{-0.021}^{+0.021}$  & Stellar radius ($R_\odot$) & \texttt{isochrones} fit (\S\ref{sec:isochrones}) \\
$M_\star$ &  $1.150_{-0.071}^{+0.072}$ & Stellar mass ($M_\odot$) & \texttt{isochrones} fit \\
$T_{\mathrm{eff}}$ & $5873_{-49}^{+63}$ & Effective temperature (K) & \texttt{isochrones} fit \\
$\log g$ & $4.270_{-0.022}^{+0.021}$ & Surface gravity (cgs)& \texttt{isochrones} fit \\

[Fe/H] & $0.230_{-0.053}^{+0.060}$ & Metallicity (dex) & \texttt{isochrones} fit  \\
$\tau_\star$ & $4.79_{-2.90}^{+2.85}$ & Stellar age (Gyr) & \texttt{isochrones} fit  
\end{tabular}
\end{table*}

\subsection{Spectroscopic Analysis}\label{sec:spec}

The high-resolution spectra from PFS allowed the bulk characteristics of TIC-186936449 to be measured with high precision.  We used the \texttt{SPECIES}\footnote{https://github.com/msotov/SPECIES} code \citep[][]{soto2018spectroscopic,soto2021spectroscopic} to calculate the first-pass effective temperature, $\log g$, and metallicity.  \texttt{SPECIES} principally operates by measuring the equivalent widths (EWs) of a number of unblended Fe~\sc{i}\rm~and Fe~\sc{ii}\rm~lines in input iodine-free spectra like that from PFS, making use of the radiative transfer code \texttt{MOOG} \citep{sneden1973nitrogen}.  The initialisation points input to \texttt{MOOG} to begin the analysis of the iron line EWs came from the prior photometric measurements listed in Table~\ref{tab:StellarParams}.  \texttt{ATLAS9} model atmospheres \citep{kurucz1993atlas9} provide the basis structure models for \texttt{MOOG} to solve the radiative transfer equation.  As part of the process, any correlation between the Fe abundances and the reduced EWs (EW/$\lambda$) or species excitation potentials are minimised using an iterative procedure that runs up to a maximum of 10,000 iterations, and all computations were performed assuming local thermodynamical equilibrium (LTE). The final main parameters from \texttt{SPECIES} are: $T_{\rm eff} = 5864 \pm 50$\,K, $\text{[Fe/H]} = 0.23 \pm 0.05$\,dex, $\log \text{g} = 4.17 \pm 0.09$, and $v\sin{i} = 5.00 \pm 0.28$\, km/s.

\subsection{SED Analysis}\label{sec:sed}

The second-pass analysis to further constrain the stellar parameters, whilst providing an additional analysis incorporating the Gaia spectrophotometry check of the \texttt{SPECIES} results, was performed using the \texttt{ARIADNE}\footnote{https://github.com/jvines/astroariadne} code \citep{vines2022ariadne}.  \texttt{ARIADNE} can either run independently of a previously performed spectral analysis, or make use of the spectral results as input priors to further refine the modeling.  In this case we performed both, however we only report the results where we used the input \texttt{SPECIES} measurements since the precisely measured $\log g$ and metallicity, in particular, help to better constrain the other bulk parameters. \texttt{ARIADNE} makes use of a custom designed Bayesian Model Averaging (BMA) spectral energy distribution (SED) fitting method to precisely measure the $T_{\mathrm{eff}}$ and $R_\star$.  The SED is constrained by the input photometric measurements from various ground- and space-based programs, shown in Table~\ref{tab:StellarParams}.  The appropriate SED models are Phoenix V2 \citep{husser2013new}, BT-Settl \citep{hauschildt1999nextgen,allard2012models}, ATLAS9 \citet{castelli2004new}, and Kurucz \citet{kurucz1993atlas9}. \texttt{ARIADNE} results are as follows, T$_{\rm eff} = 5822^{+66}_{-54}$\,K, $\log \text{g} = 4.3 \pm 0.4$, $R_{\star} = 1.31 \pm 0.05$\,R$_{\sun}$ and $M_{\star} = 1.16 \pm 0.07$\,M$_{\sun}$.

\begin{figure*}
    \centering
    \includegraphics[width=0.67\linewidth,trim=80 70 50 50,clip]{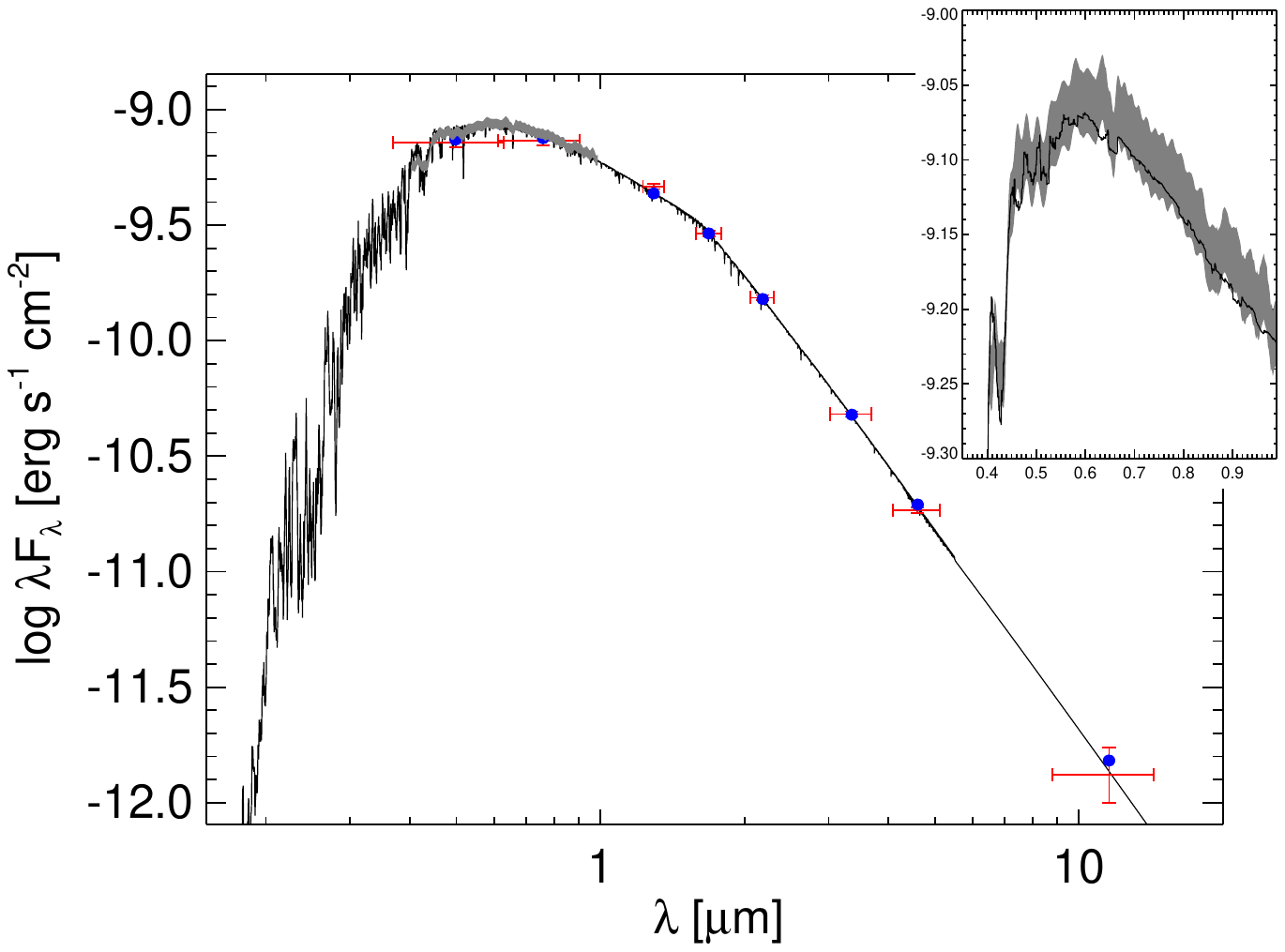}
\caption{Spectral energy distribution of TOI-2665. Red symbols represent the observed photometric measurements, where the horizontal bars represent the effective width of the passband. Blue symbols are the model fluxes from the best-fit PHOENIX atmosphere model (black). The gray swathe (also in the inset) represents the absolute flux-calibrated spectrophotometry from {\it Gaia}. \label{fig:sed}}
\end{figure*}

As a further independent determination of the basic stellar parameters, we performed an analysis of the broadband SED analysis of the star together with the {\it Gaia\/} DR3 parallax \citep[with no systematic offset applied; see, e.g.,][]{StassunTorres:2021}, in order to determine an empirical measurement of the stellar radius, following the procedures described in \citet{Stassun:2016,Stassun:2017,Stassun:2018}. We pulled the $JHK_S$ magnitudes from {\it 2MASS}, the $G_{\rm BP} G_{\rm RP}$ magnitudes from {\it Gaia}, and the W1--W3 magnitudes from {\it WISE}. 
We also utilized the absolute flux-calibrated spectrophotometry from {\it Gaia}. 
Together, the available photometry spans the full stellar SED over the wavelength range 0.4--20~$\mu$m (see Figure~\ref{fig:sed}).  

We performed a fit using PHOENIX stellar atmosphere models \citep{Husser:2013}, with the effective temperature ($T_{\rm eff}$), surface gravity ($\log g$), and metallicity ([Fe/H]) adopted from the spectroscopic analysis. The extinction, $A_V$, was limited to maximum line-of-sight value from the Galactic dust maps of \citet{Schlegel:1998}. The resulting fit (Figure~\ref{fig:sed}) has a reduced $\chi^2$ of 1.8, with a best-fit $A_V = 0.06 \pm 0.05$. Integrating the (unreddened) model SED gives the bolometric flux at Earth, $F_{\rm bol} = 1.183 \pm 0.028 \times 10^{-9}$ erg~s$^{-1}$~cm$^{-2}$. Taking the $F_{\rm bol}$ together with the {\it Gaia\/} parallax directly gives the bolometric luminosity, $L_{\rm bol} = 1.834 \pm 0.043$~L$_\odot$. The stellar radius follows from the Stefan-Boltzmann relation, giving $R_\star = 1.312 \pm 0.032$~R$_\odot$. In addition, we can estimate the stellar mass from the empirical relations of \citet{Torres:2010}, giving $M_\star = 1.07 \pm 0.06$~M$_\odot$. 

Finally, we can also estimate the projected stellar rotation period from the spectroscopic $v\sin i$ together with $R_\star$, giving $P_{\rm rot} /\sin i = 15.2 \pm 0.9$ days, from which we can estimate an upper limit on the system age via the empirical gyrochronology relations of \citet{Mamajek:2008}, giving $1.6 \pm 0.2$~Gyr.

\subsection{Isochrones Analysis}\label{sec:isochrones}
To derive a final self-consistent set of physical parameters for the host star, we fit the observed properties of TOI-2665 to MIST evolutionary models \citep{Dotter2016, Choi2016} using the stellar model grid package \texttt{isochrones} \citep{Morton2015}. We defined a single-star model using parallax from Gaia DR3 \citep{Gaia,GaiaDR3}, $JHK_{s}$ magnitudes from 2MASS \citep{2MASS}, $GG_{\mathrm{BP}}G_{\mathrm{RP}}$ magnitudes from Gaia DR3, $W1-W3$ magnitudes from WISE \citep{WISE}, and $T_{\mathrm{eff}}$ and [Fe/H] from the PFS spectral analysis. Following the convention of \citet{Eastman2019}, we inflated the Gaia magnitude uncertainties to 0.02.

We fit for stellar mass, radius, metallicity, distance, and age, and used the default priors defined by \texttt{isochrones} with the exception of a broad flat prior for metallicity ([Fe/H] $\in [0, 2]$). The fit was run using an $\texttt{emcee}$ \citep{emcee} ensemble sampler with 100 walkers over 50,000 iterations, and considered the fit converged based on the chain being more than 50 times longer than the estimated autocorrelation time. After discarding the first 1000 steps as burn-in, we find that TOI-2665 is a high-metallicity G dwarf ([Fe/H] $= 0.23_{-0.05}^{+0.06}$ dex, $\log g = 4.27_{-0.02}^{+0.02}$, $T_{\mathrm{eff}} = 5873_{-49}^{+63}$ K) with mass $M_\star=1.15_{-0.03}^{+0.03}~M_\odot$ and radius $R_\star = 1.30_{-0.01}^{+0.01}~R_\odot$. All properties are within $1\sigma$ of those derived from both the PFS and SED analyses.

The \texttt{isochrones} parameter uncertainties are underestimated because they do not take into account the systematic error arising from differences between stellar grid models. Following \citet{Tayar2022}, we use the \texttt{kiauhoku} pacakge \citep{Claytor2020} to estimate the properties of TOI-2665 based on interpolating effective temperature, luminosity, and metallicity to stellar grid models from YREC \citep{Demarque2008}, MIST, DSEP \citep{Dotter2008}, and GARSTEC \citep{WeissSchlattl2008} codes and databases. We find maximum differences in stellar mass, radius, log$g$, and age at $0.065~M_{\odot}$, $0.018~R_{\odot}$, $0.016$, and $2.70$ Gyr, respectively, between different grid solutions. These systematic errors have been added in quadrature to the fit uncertainties and are reflected in Table \ref{tab:StellarParams}. We adopt these final parameters when requiring stellar properties in the remainder of this paper.

\section{Planet Parameters}\label{sec:planet}

\subsection{Joint transit +  RV model fit}

We performed a joint fit of our multi-sector TESS light curve and ground-based Brierfield light curve with the Magellan II/PFS RV measurements using PyMC \citep{pymc2023}, a Hamiltonian Monte Carlo posterior sampler within the \texttt{exoplanet} package \citep{exoplanet}. The planet parameters were orbital period ($P$), reference transit epoch ($t_{0}$), log of the planet-to-star radius ratio ($\log(R_{p}/R_{\star})$), impact parameter ($b$), as well as the RV semi-amplitude ($K$), orbital eccentricity ($e$), and argument of pericenter ($\omega$). All parameters were assigned uniform priors, with the exception of $e$ and $\omega$ with priors following \citet{2013Kippingeccentricity} as implemented in \texttt{exoplanet}. We further fit for quadratic limb-darkening coefficients ($u_{1}$, $u_{2}$) parameterized by \citet{Kipping203limbdarkening} for both the TESS band and Brierfield $R_{c}$ band, mean flux and jitter terms for each TESS sector and the ground-based light curve, and RV zero-point offset (a small correction to the baseline velocity) and jitter terms, with uniform priors. The model was sampled with four chains with 1000 tuning and 1000 drawing steps each, and convergence was confirmed based on the Gelman-Rubin convergence statistic of each parameter satisfying $r < 1.01$. Corner plots of the resulting fits are shown in Figure \ref{figure: Corner plots for join fit}. Because we fixed the stellar parameters ($R_{\star}$, $M_{\star}$) to the median \texttt{isochrones}-fitted results, we propagated their uncertainties into the physical planet parameters derived from the fits.

\begin{figure*}[t]
    \centering
    \includegraphics[width=0.48\linewidth]{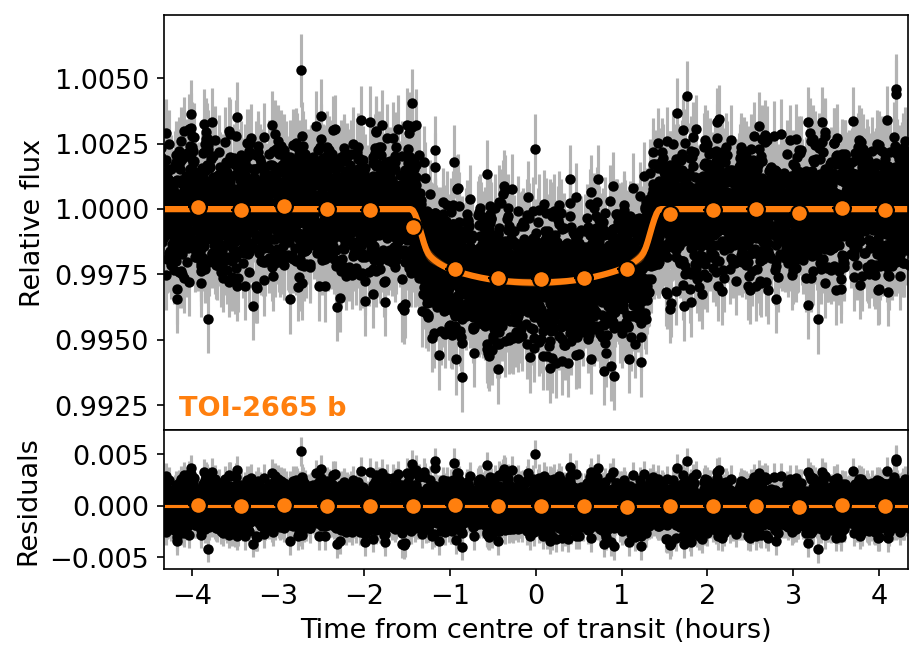}
    \includegraphics[width=0.46\linewidth]{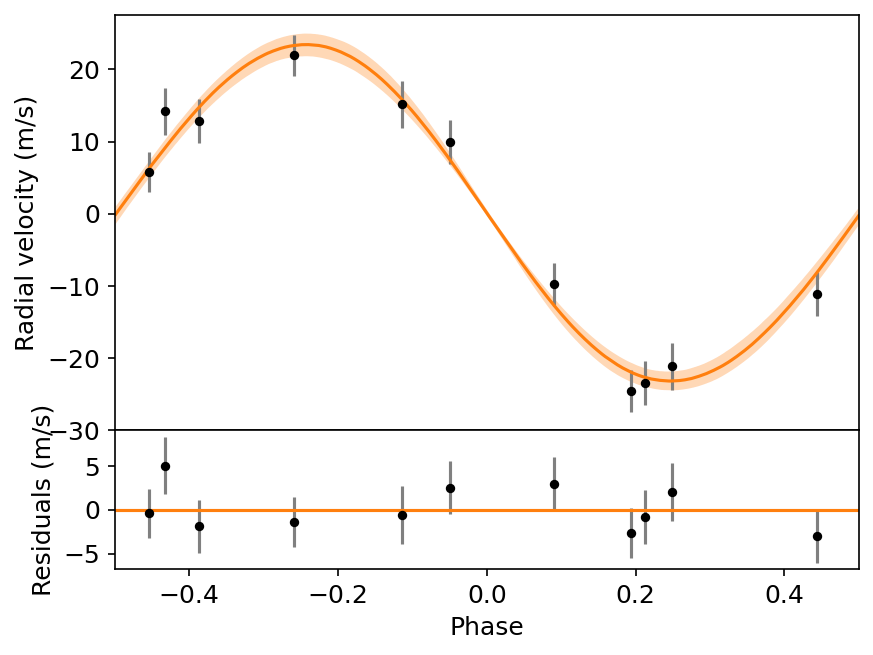}
    \caption{The phase-folded TESS and ground-based Brierfield light curves (left) and PFS radial velocity measurements (right). Black points show the underlying data with uncertainties in grey, while the orange points in the transit phase diagram represent data binned every 30 minutes. Orange lines represent our median model values following the joint transit-RV fit, while the shaded  regions cover 16th to 84th percentiles. The residuals show the data after removal of the median model.}
    \label{fig:TransitFit}
\end{figure*}

The model fit results are provided in Table \ref{tab:PlanetaryParams}, and Figure \ref{fig:TransitFit} shows the phase-folded TESS light curves and PFS RV observations. We measure a mass of $M_{p} =  57.0_{-4.2}^{+4.2} ~M_{\oplus}$ for TOI-2665.01, which is a $13.5\sigma$ recovery of the planet. Our joint transit and RV fit therefore confirms the planetary nature of this object, which we now refer to as TOI-2665 b.

\begin{figure}[t]
    \centering
    \includegraphics[width=1\linewidth]{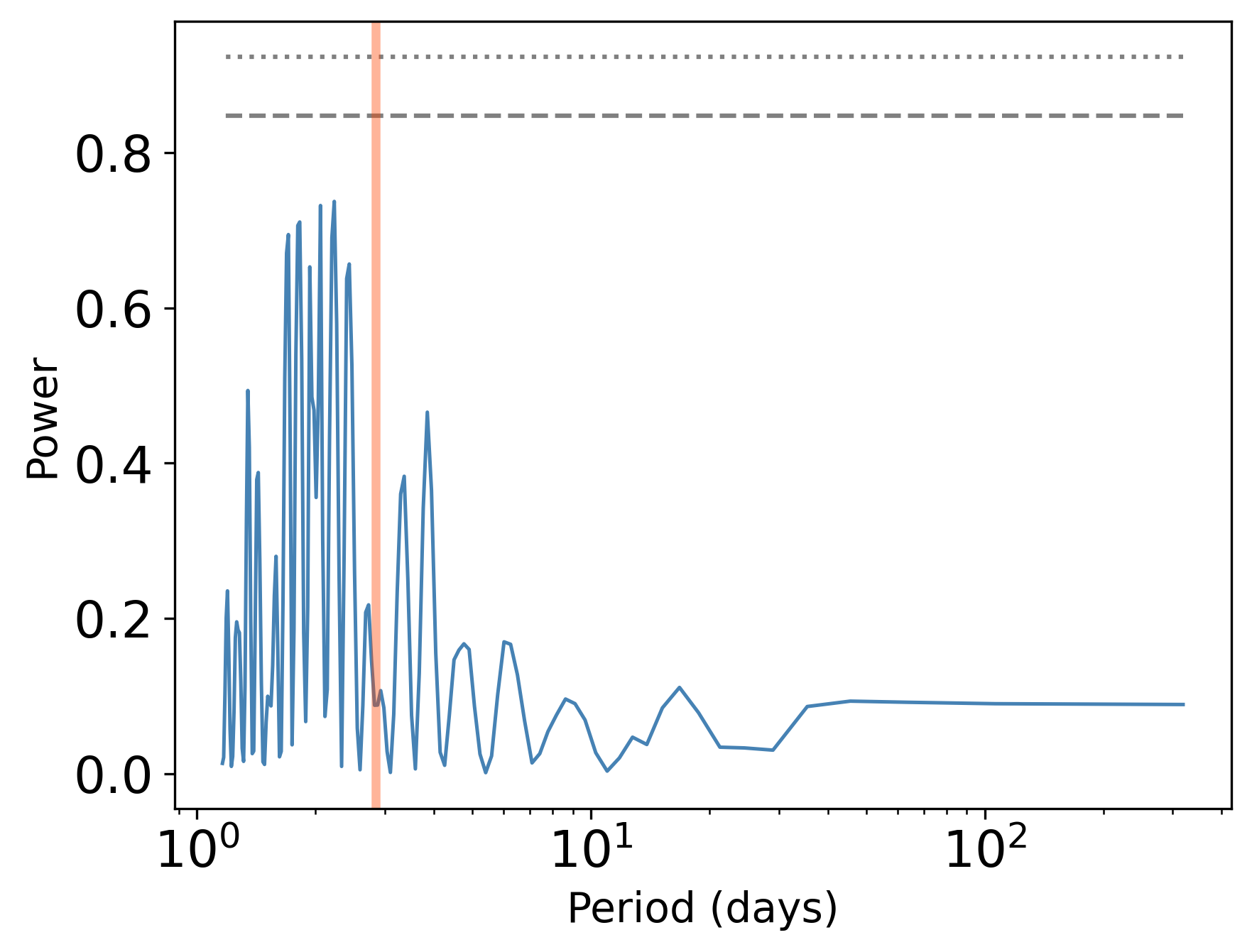}
    \caption{GLS periodogram for PFS RV measurements after subtracting the fitted model for TOI-2665 b. The shaded orange line corresponds to the planet's period of 2.86 days. The dashed and dotted lines correspond to 0.1 and 0.01\% false alarm probability levels. There are no significant peaks in the periodogram, suggesting that there are no additional planets detectable in the data.}
    \label{fig:GLSperiodogram}
\end{figure}

While the TESS light curves do not display additional transits, we searched for evidence of companions in the RVs. First, we ran a GLS periodogram on the RV residuals, and found no statistically significant peaks (Figure \ref{fig:GLSperiodogram}). Second, we re-fit our joint transit+RV model with an additional linear slope to model the RV signals from a potential long-period companion, finding a slope that was consistent with being flat (slope = $0.06\pm0.08$). We conclude that TOI-2665 b is likely a single planet, in line with typical hot Neptune desert-dwellers \citep{Dong2018,VissapragadaBehmard2025}. 

\begin{table*}[]
    \centering
    \caption{Joint Transit-RV Fit Results for TOI-2665 b.}\label{tab:PlanetaryParams}
\begin{tabular}{lll}
\hline\hline
Parameter & Value & Description \\
\hline
\multicolumn{3}{c}{\emph{Fitted Planet Parameters}} \\
\hline
$P$ (days) & $2.845650 ^{+0.000001}_{-0.000001}$ & Orbital Period in days \\
$t_0$ (BJTD) & $1518.6446 ^{+0.0007}_{-0.0007}$  & Transit epoch (BJD$\mathrm{_{TDB}}$ - 2457000) \\
$\log({R_p/ R_\star})$ & $-3.01 ^{+0.01}_{-0.02}$  & Log of the planet-to-star radius ratio  \\
$b$ & $0.526^{+0.040}_{-0.082}$ & Impact parameter \\
$K$ (m s${^{-1}}$) & $23.42^{+1.33}_{-1.42}$ & RV semi-amplitude (m s$^{-1}$) \\
$e$ & $0.03^{+0.04}_{-0.02}$ & Orbital eccentricity  \\
$\omega$ & $51^{+57}_{-130}$ & Argument of pericenter ($^\circ$) \\
\hline
\multicolumn{3}{c}{\emph{Derived Planet Properties} }\\
\hline
$R_p$ & $7.00^{+0.16}_{-0.15}$ & Planet radius ($R_\oplus$) \\
$M_p$ & $57.0^{+4.2}_{-4.2} $ & Planet mass ($M_\oplus$) \\
$\rho_p$ & $0.913 ^{+0.090}_{-0.086}$ & Planet density (g cm$^{-3}$) \\
$a$ & $0.0412^{+0.0009}_{-0.0009}$ & Semi-major axis (au) \\
$S$ & $1068^{+68}_{-64}$ & Insolation Flux ($S_\oplus$) \\
$T_{\mathrm{eq}}$ & $1593^{-24}_{+25}$ & Equilibrium temperature\footnote{Assuming albedo of 0} ($K$) \\
$i$ & $85.54^{+0.51}_{-0.26}$ & Inclination angle ($^\circ$)\\
$\tau$ & $2.892^{+0.084}_{-0.080}$ & Transit duration (hrs) \\
$R_{p}/R_{\star}$ & $0.0494_{-0.0007}^{+0.0006}$ & Planet-to-star radius ratio \\
$a/R_{\star}$ & $6.79_{-0.18}^{+0.18}$ & Planet separation in units of stellar radius \\
\hline
\multicolumn{3}{c}{\emph{Fitted Photometric Parameters}} \\
\hline
$\mu_{\mathrm{Sec8}}$ & $-9^{+33}_{-33}$ & Sector 8 flux offset (ppm) \\
$\mu_{\mathrm{Sec35}}$ & $13^{+33}_{-32}$ & Sector 35 flux offset (ppm) \\
$\mu_{\mathrm{Sec62}}$ & $75^{+27}_{-27}$ & Sector 62 flux offset (ppm) \\
$\mu_{\mathrm{Sec89}}$ & $5^{+24}_{-24}$ & Sector 89 flux offset (ppm) \\
$\mu_{\mathrm{Brierfield}}$ & $113^{+156}_{-161}$ & Brierfield flux offset (ppm) \\

$\sigma_{\mathrm{Sec8}}$ & $94^{+63}_{-60}$ & Sector 8 flux jitter (ppm) \\
$\sigma_{\mathrm{Sec35}}$ & $80^{+73}_{-54}$ & Sector 35 flux jitter (ppm) \\
$\sigma_{\mathrm{Sec62}}$ & $116^{+99}_{-80}$ & Sector 62 flux jitter (ppm) \\
$\sigma_{\mathrm{Sec89}}$ & $78^{+82}_{-54}$ & Sector 89 flux jitter (ppm) \\
$\sigma_{\mathrm{Brierfield}}$ & $141^{+147}_{-99}$ & Brierfield flux jitter (ppm) \\
$u_{1,\mathrm{TESS}}$ & $0.37^{+0.22}_{-0.22}$ & TESS band quadratic limb-darkening parameter 1 \\
$u_{2,\mathrm{TESS}}$ & $0.20^{+0.35}_{-0.33}$ & TESS band quadratic limb-darkening parameter 2 \\
$u_{1,\mathrm{Brierfield}}$ & $0.51^{+0.44}_{-0.34}$ & Brierfield $R_{c}$ band quadratic limb-darkening parameter 1 \\
$u_{2,\mathrm{Brierfield}}$ & $0.10^{+0.41}_{-0.41}$ & Brierfield $R_{c}$ band quadratic limb-darkening parameter 2 \\
\hline
\multicolumn{3}{c}{\emph{Fitted RV Parameters}} \\
\hline
$\mu_{\mathrm{PFS}}$ & $-4.78^{+0.96}_{-0.93}$ & PFS RV zero-point offset (m s$^{-1}$)  \\
$\sigma_{\mathrm{PFS}}$ & $2.51^{+1.29}_{-0.95}$ & PFS RV jitter (m s$^{-1}$) \\
\end{tabular}
\end{table*}

\section{Discussion}\label{sec:discussion}

\begin{figure*}[t]
    \centering
    \includegraphics[width=0.48\linewidth]{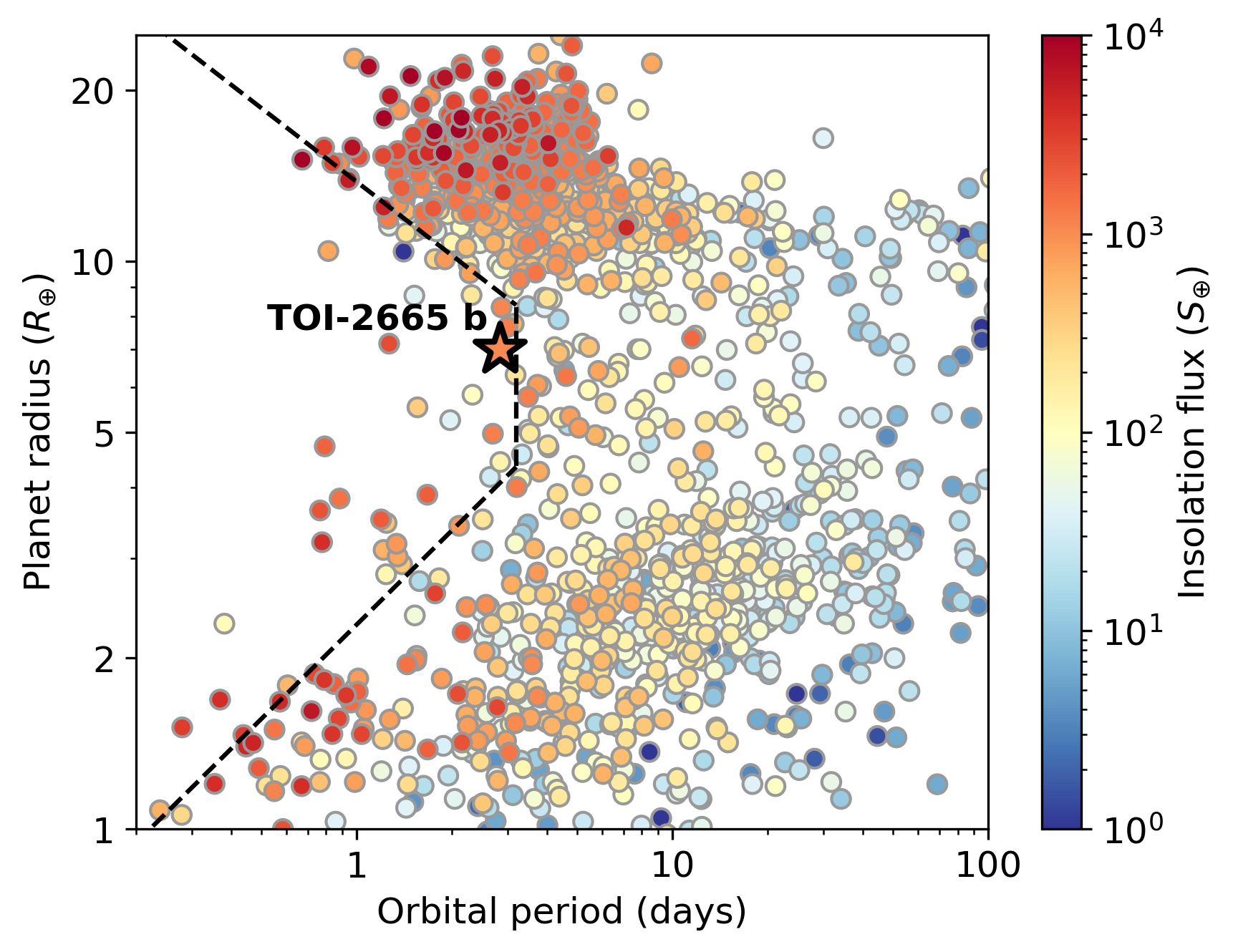}
    \includegraphics[width=0.48\linewidth]{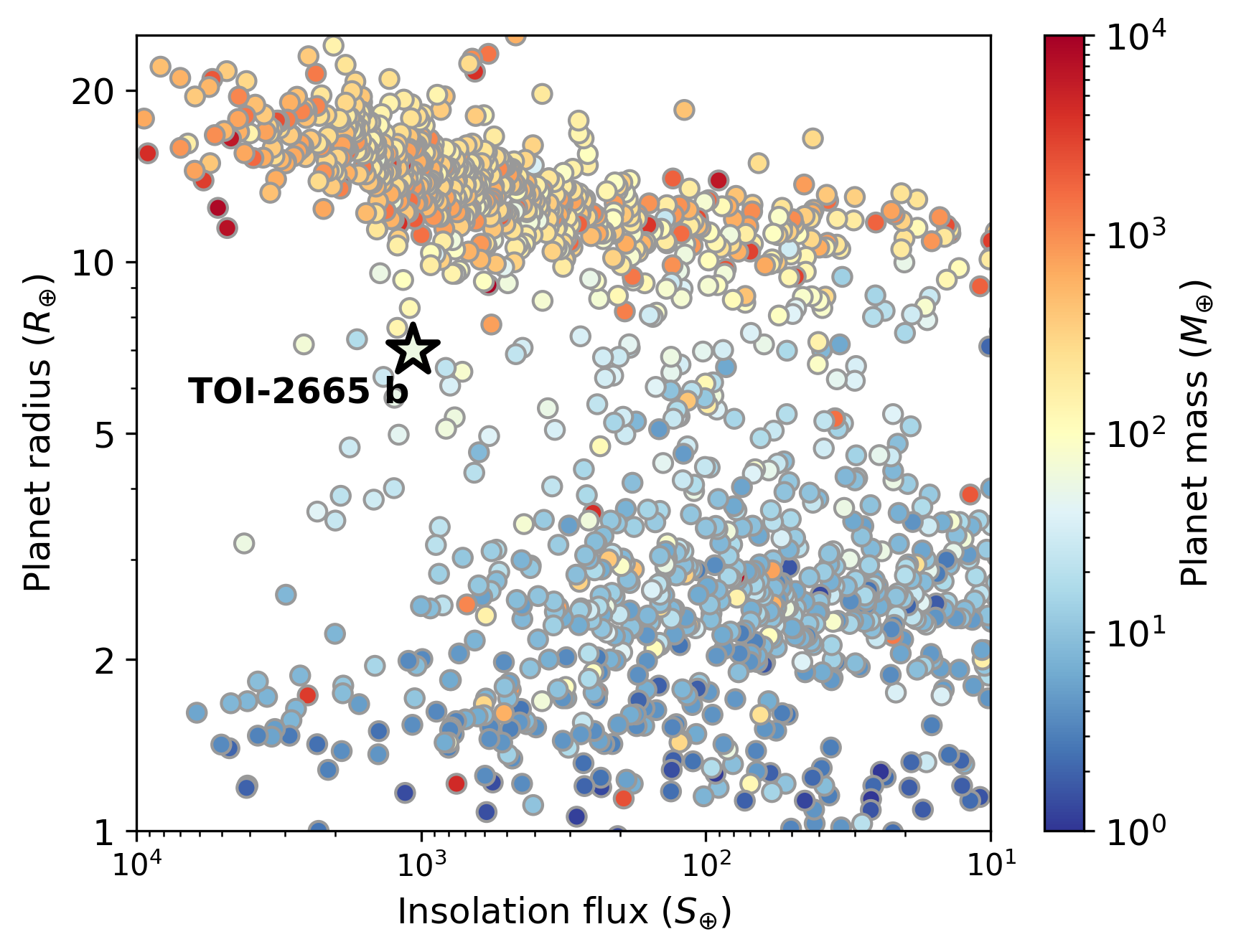}
    \caption{Planets with mass measurements from the NASA Exoplanet Archive's Planetary Systems Composite Data table \citep{NEA}, accessed 2026 March 25, showing radius vs. orbital period (left) and insolation flux (right). TOI-2665 b is plotted as the black star. The black dashed lines indicate the radius-period boundaries of the hot Neptune desert from \citet{Castro-Gonz_NpetuneDesertLines}.} 
    \label{fig:desert}
\end{figure*}

\begin{figure*}
    \centering
    \includegraphics[width=0.48\linewidth]{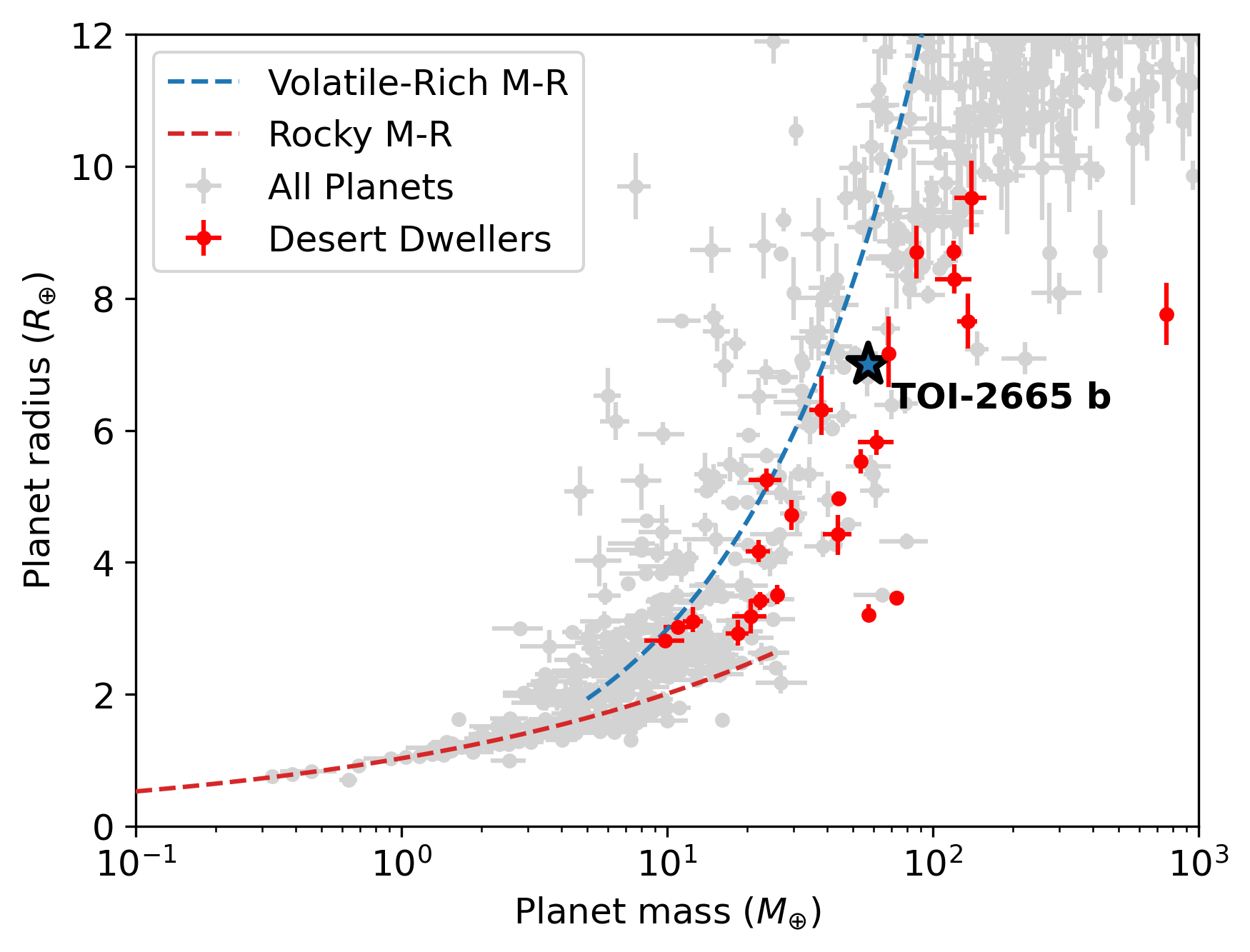}
    \includegraphics[width=0.49\linewidth]{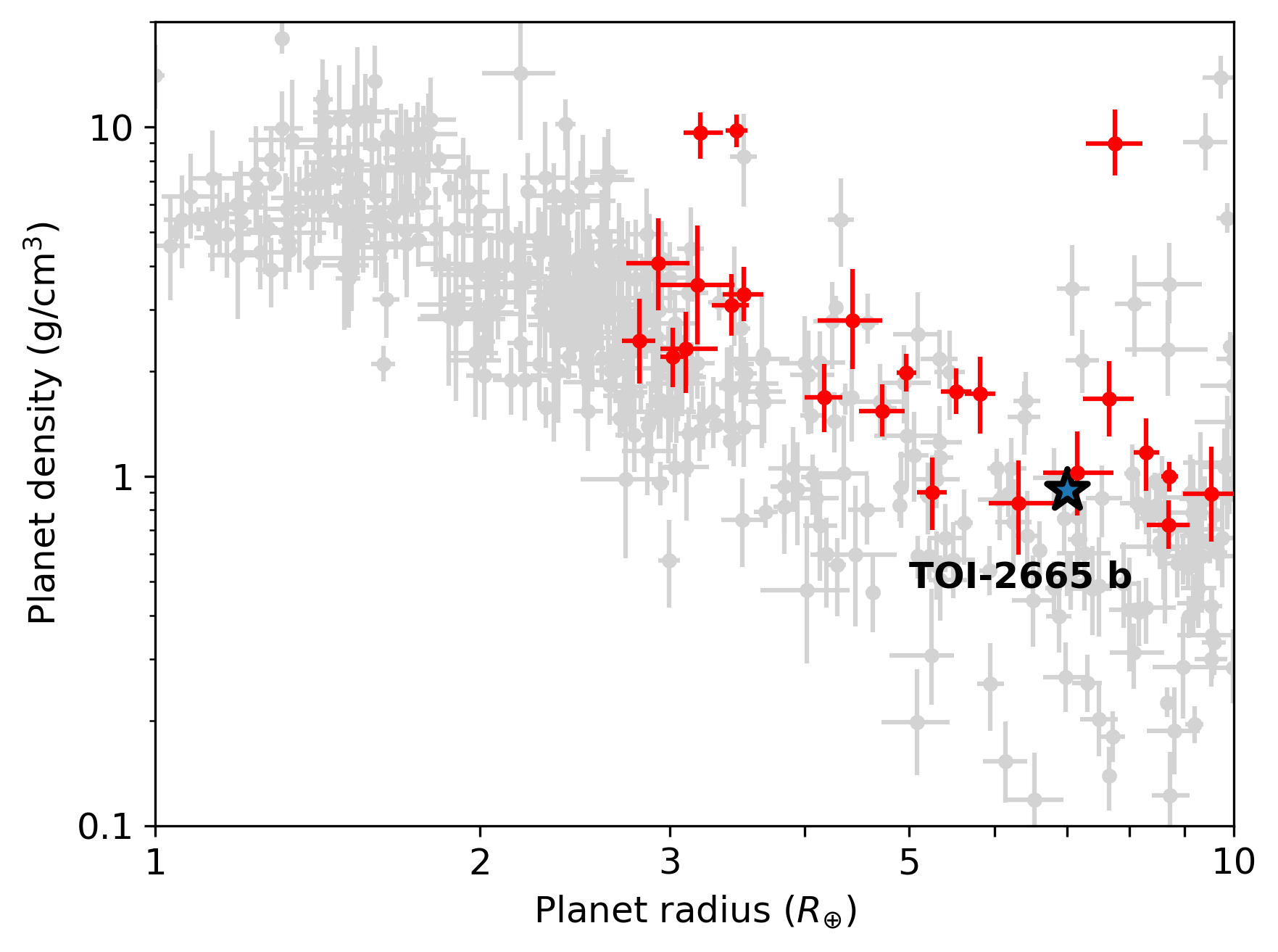}
    \caption{Same as Figure \ref{fig:desert}, but for mass vs. radius (left) and density vs. radius (right), and only showing planets with masses and radii measured to better than 25\% and 10\% precision, respectively. Planets within the Neptune desert boundaries defined by \citet{Castro-Gonz_NpetuneDesertLines} and with $R_{p} \in [2, 10]~R_{\oplus}$ are shown in red. The volatile-rich and rocky planet mass-radius relations from \citet{Otegi2020} are also shown for comparison. Like other desert-dwellers, TOI-2665 b is somewhat higher density than typical volatile-rich planets at similar sizes.}\label{fig:mass_radius}
\end{figure*}

With an orbital period $P = 2.846$ days, radius $R_{p} = 7.00_{-0.15}^{+0.16}~R_{\oplus}$, and mass $M_{p} = 57.0_{-4.2}^{+4.2}~M_{\oplus}$, TOI-2665 b joins the small list of eleven sub-Saturns ($4 < R_{p} < 8~R_{\oplus}$) found within the hot Neptune desert \cite[$P \lesssim 3.2$ days for this size of planet;][]{Castro-Gonz_NpetuneDesertLines} (left panel of Figure \ref{fig:desert}). Like other desert-dwellers, the density of TOI-2665 b ($\rho \sim 0.9$ g cm$^{-3}$) is relatively high compared to other planets of its size (Figure \ref{fig:mass_radius}), implying an enriched heavy-element mass fraction. Among desert-dwellers in mass-radius space, TOI-2665 b is most similar to HATS-7 b \citep{Bakos2015}, TOI-532 b \citep{Kanodia2021}, TOI-3072 b \citep{TOI3071b}, and TOI-3862 b \citep{Carleo2026}, all of which also orbit metal-rich stars ([Fe/H] = $0.1 - 0.38$ dex). TOI-2665 b especially stands out in insolation flux space (right panel of Figure \ref{fig:desert}), where it is significantly more irradiated ($S = 1068~S_{\oplus}$) than most other planets of its size.

\subsection{Internal Structure and Evaporation History}\label{sec:evap}

As a resident of the hot Neptune desert, TOI-2665 b may represent either a planet actively losing its atmosphere or one that has retained a substantial envelope despite prolonged exposure to intense stellar irradiation. Modeling its atmospheric evolution therefore provides a means of assessing whether or not its survival in the desert requires an unusually resilient atmosphere.

We used the \texttt{photoevolver} codebase\footnote{\url{https://github.com/jorgefz/photoevolver}} to evolve the atmosphere of TOI-2665 b, taking into account the planet's internal structure and cooling rate, atmospheric mass loss processes, and the stellar emission history \citep{Fernandez2023}. \texttt{photoevolver} assumes a two-layer internal structure involving a rocky core (with mass $M_{\mathrm{core}}$ and radius $R_{\mathrm{core}}$) surrounded by a pure H/He envelope (with mass $M_{\mathrm{env}}$ and thickness $R_{\mathrm{env}}$). Following \citet{TOI3071b}, who modeled the evolution of two desert dwellers with similar properties as TOI-2665 b (TOI-2374 b and TOI-3071 b), we adopted the empirical mass-radius relation by \citet{Otegi2020} for the rocky core, the envelope structure model by \citet{Chen2016} for the gaseous atmosphere, and the stellar rotation and XUV evolution model presented by \citet{Johnstone2021}. The two-layer internal structure model predicts a present-day high core (heavy-element) mass of $M_{\mathrm{core}} = 38.7\pm3.1~M_{\oplus}$, corresponding to an atmospheric mass fraction of $f_{\mathrm{env}} \equiv M_{\mathrm{env}}/M_{p} = 32.1\pm7.6\%$.

Setting the stellar rotation period to $P_{\mathrm{rot}} = 15.2$ days and age to $\tau_{\star} = 1.6$ Gyr (assuming the star's rotation is aligned with our line of sight, i.e., setting $\sin{i} = 1$ in our analysis from \S\ref{sec:sed}), we simulated the evolution of the planet starting at 10 Myr, with a variable time step of 1 Myr. For the mass-loss mechanism, we adopted the hydrodynamic models and interpolator described by \citet{Kubyshkina2018}. For our nominal results, we adopted the median radius and mass for TOI-2665 b. For a more extreme evaporation history, we adopted the $+1\sigma$ value for radius and $-1\sigma$ value for mass, which reduces planet density and therefore increases escape rates. For a less extreme history, we picked the opposite parameters.

\begin{figure}
    \includegraphics[width=\linewidth]{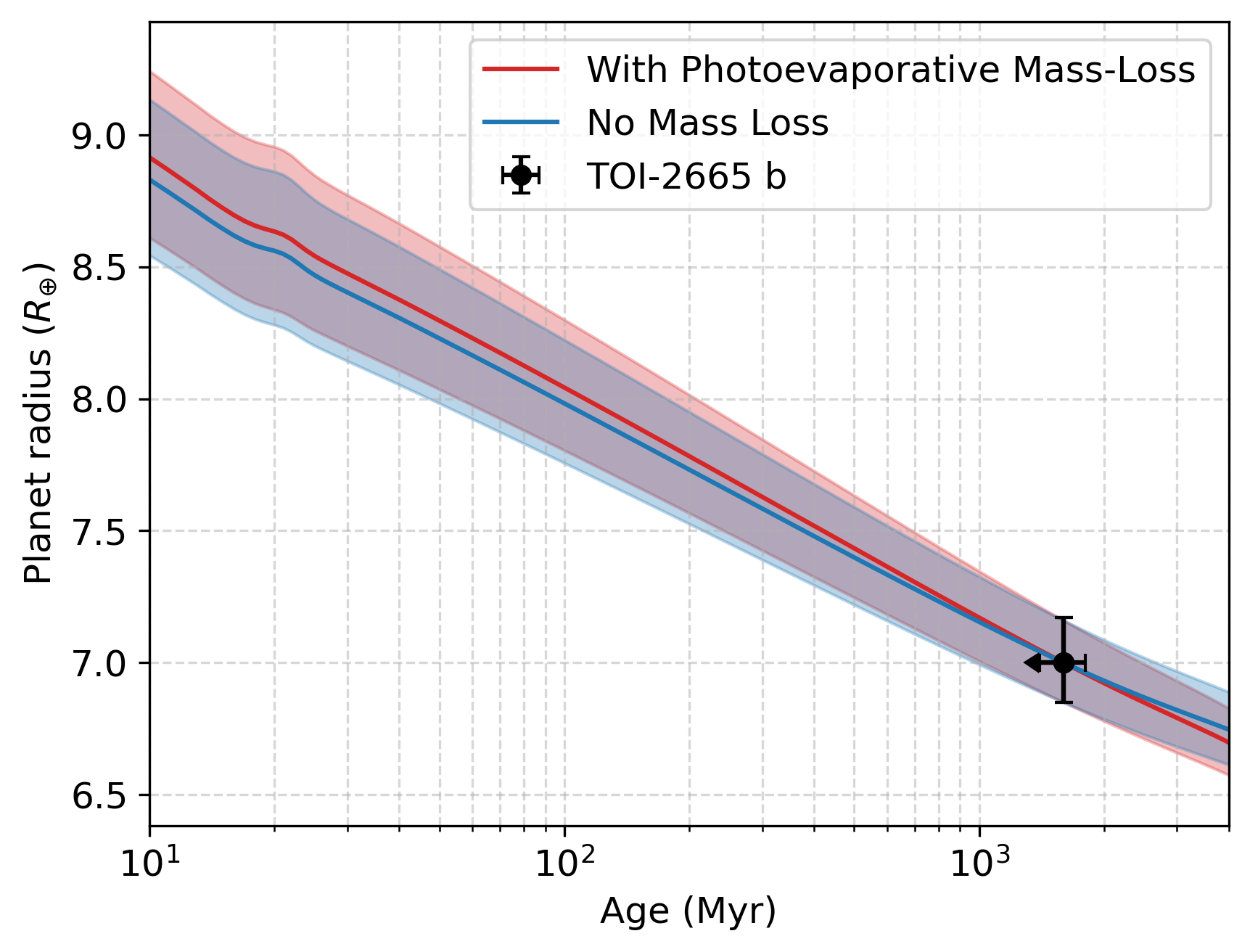}
    \caption{The evolution of the radius of TOI-2665 b under photoevaporation (red) and under no atmospheric mass-loss (blue), assuming a pure H/He envelope, a present-day age of $\tau_{\star} = 1.6$ Gyr (an upper limit; see \ref{sec:sed}), and a constant orbital period of $P = 2.85$ days. The shaded areas show possible evaporation histories taking into account the uncertainties in the planet's mass and radius. Overall, TOI-2665 b is stable against photoevaporation, and the evolution of its radius is largely driven by thermal contraction (assuming no disruptive, non-evaporative mass-loss processes have occurred).}\label{fig:evolution}
\end{figure}

The change in the planet's radius over time is shown in Figure \ref{fig:evolution}, demonstrating similar results to a history assuming no atmospheric mass-loss. We find that the atmospheric mass fraction ($f_{\mathrm{env}} \equiv M_{\mathrm{env}}/M_{p}$) has dropped by only $\sim0.6\%$, from $\sim32.7\%$ to $\sim32.1\%$ over the course of its lifetime. These estimated evaporative mass-loss rates should also be regarded as upper limits, as the simulations assume the planet has persisted in its $P = 2.85$-day orbit (rather than migrated inward from a less irradiated location), and assume a pure H/He atmosphere. Given the metal-rich nature of the host star, the planet's atmosphere is likely enriched in heavier elements, increasing its mean molecular weight and reducing the efficiency of hydrodynamic escape \citep{OwenJackson2012, OwenMurray-Clay2018.480.2206O}. Assuming the same orbit, we find that $f_{\mathrm{env}}$ should only drop a further $\sim1.7\%$ through TOI-2665's 6.6 Gyr main sequence lifetime. Overall, we find that a high-metallicity atmosphere is not required to explain the survival of the planet in the hot Neptune desert.

\subsection{Possible Origins in the Context of the Neptune Desert}

TOI-2665 b may have initially formed as a volatile-rich sub-Saturn-sized planet, in which case its evaporative history suggests that its radius evolution has been dominated by thermal contraction, with atmospheric escape contributing only a minor correction. In this case, some metal enrichment is needed to explain its relatively high density compared to other sub-Saturns, especially given that its high heavy-element mass of $\sim40~M_{\oplus}$ (\S\ref{sec:evap}) is a challenge to explain with standard formation models \cite[e.g.][]{Pollack1996,Bitsch2018}. Potential solutions could include a ``bottom up'' formation scenario where TOI-2665 b formed via planetary mergers \citep{Ogihara2021,Ginzburg2020}, or post-formation enrichment due to accretion of planetesimals \citep{Mousis2009,Shibata2020}.

Alternatively, the planet's history could have involved significantly more atmospheric mass-loss due to non-evaporative processes, such that TOI-2665 b represents the remnant core of a once more massive giant. This ``top down'' pathway has been proposed as the most common origin for hot Neptune desert-dwellers, which follow the same host star metallicity distribution as hot Jupiters, unlike cooler Neptunes (\citet{VissapragadaBehmard2025}; see also \citet{Dong2018} for an earlier work noting the similarities between hot Jupiters and hot Neptunes). Fittingly, the high stellar metallicity of TOI-2665 ([Fe/H] = $0.23_{-0.05}^{+0.06}$ dex) is similar to the metallicities of typical hot Jupiter hosts (mean [M/H] = 0.22 dex; \citet{VissapragadaBehmard2025}). In terms of the subsequent mass-loss mechanism, Roche-lobe overflow (RLO) has frequently been shown to be able to remove the atmospheres of hot Jupiters, at least partially \cite[e.g., ][]{Valsecchi2014, Jackson2016, Thorngren2023, Lazovik2023, Weldon2026}. Recently, \citet{Hallat2026} showed that tidally-driven RLO was capable of populating the desert as long as the angular momentum transfer during the process was inefficient. An observational feature could be host stars that rotate up to an order of magnitude more rapidly than field stars; thus, TOI-2665's true rotation period may be significantly faster than $P_{\mathrm{rot}} = 15.2$ days. Envelope disruption during high-eccentricity migration is another proposed mass-loss mechanism \citep{OwenLai2018}.

Regardless of its specific origin pathway, TOI-2665 b (or its progenitor hot Jupiter) must have either formed \textit{in situ} or migrated inwards to its present-day orbit. We do not see evidence for a distant massive companion based on the high-resolution imaging, the small-scale variations in the RVs, and the relatively low Gaia DR3 Renormalized Unit Weight Error (RUWE) value \cite[0.95, consistent with a single star;][]{Lindegren2018,Lindegren2021}; therefore, migration following a high eccentricity pumped by Kozai-Lidov oscillations driven by interactions with a stellar binary is disfavored. High eccentricity migration (and tidal circularization) following planet-planet scattering, or quiescent disk migration, could each possibly explain TOI-2665 b's arrival in the desert and its present-day circular orbit.

\subsection{Potential for Atmospheric Characterization}

Atmospheric characterization could help distinguish and explore competing formation pathways for TOI-2665 b, and constrain how much of the heavy element mass is in the core versus the envelope. The ``failed hot Jupiter'' origin would also mean that the present-day atmosphere could provide a rare observational window into the deeper layers of gas giants that are otherwise inaccessible. With a transmission spectroscopy metric \cite[TSM;][]{Kempton2018} of 71, TOI-2665 b emerges as a promising target for detailed atmospheric study.

To this end, we simulated transmission spectra using \texttt{petitRADTRANS} \citep{petitRADTRANS} under various atmospheric assumptions, and then investigated what the James Webb Space Telescope (JWST) may be able to resolve via \texttt{PandEXO} \citep{pandexo}. We first defined a logarithmically spaced pressure gradient of 100 layers between $10^{-6}$ to $100$ bars, and assumed an isothermal atmosphere set to the planet's equilibrium temperature, $T_{\mathrm{eq}} = 1593~K$. We computed the planet's reference gravity from its joint-fit-inferred physical parameters ($R_{p}$ and $M_{p}$), and set the reference pressure corresponding to the measured radius to be 0.01 bar. We used \texttt{petitRADTRANS}'s chemical equilibrium model characterized by the atmospheric carbon-to-oxygen ratio (C/O) and metallicity ([Fe/H]) to generate transmission spectra. For cloud- and haze-free atmospheres we assume solar-like C/O $= 0.55$, and simulate both $\mathrm{[Fe/H]} = 0$ and $\mathrm{[Fe/H]} = 2$ to test their possibility of being differentiated by JWST. Figure 11 shows that a single transit with the NIRISS-SOSS instrument would be enough to distinguish between a solar and high-metallicity atmosphere on TOI-2665 b.

We inputted our spectra to \texttt{PandEXO} to simulate the corresponding JWST observations by the NIRISS-SOSS instruments, then ran atmospheric retrievals with \texttt{petitRADTRANS} to attempt to recover the input C/O and [Fe/H]. We find that with a single transit observed by NIRISS-SOSS, we are able to recover [Fe/H] = $-0.13\pm0.41$ (in the Solar metallicity case) and [Fe/H] = $1.44\pm0.64$ (in the high-metallicity case). In other words, only a single transit may be needed to distinguish between solar and high-metallicity atmospheres with JWST.

\begin{figure}[t]
    \centering
    \includegraphics[width=1\linewidth]
    {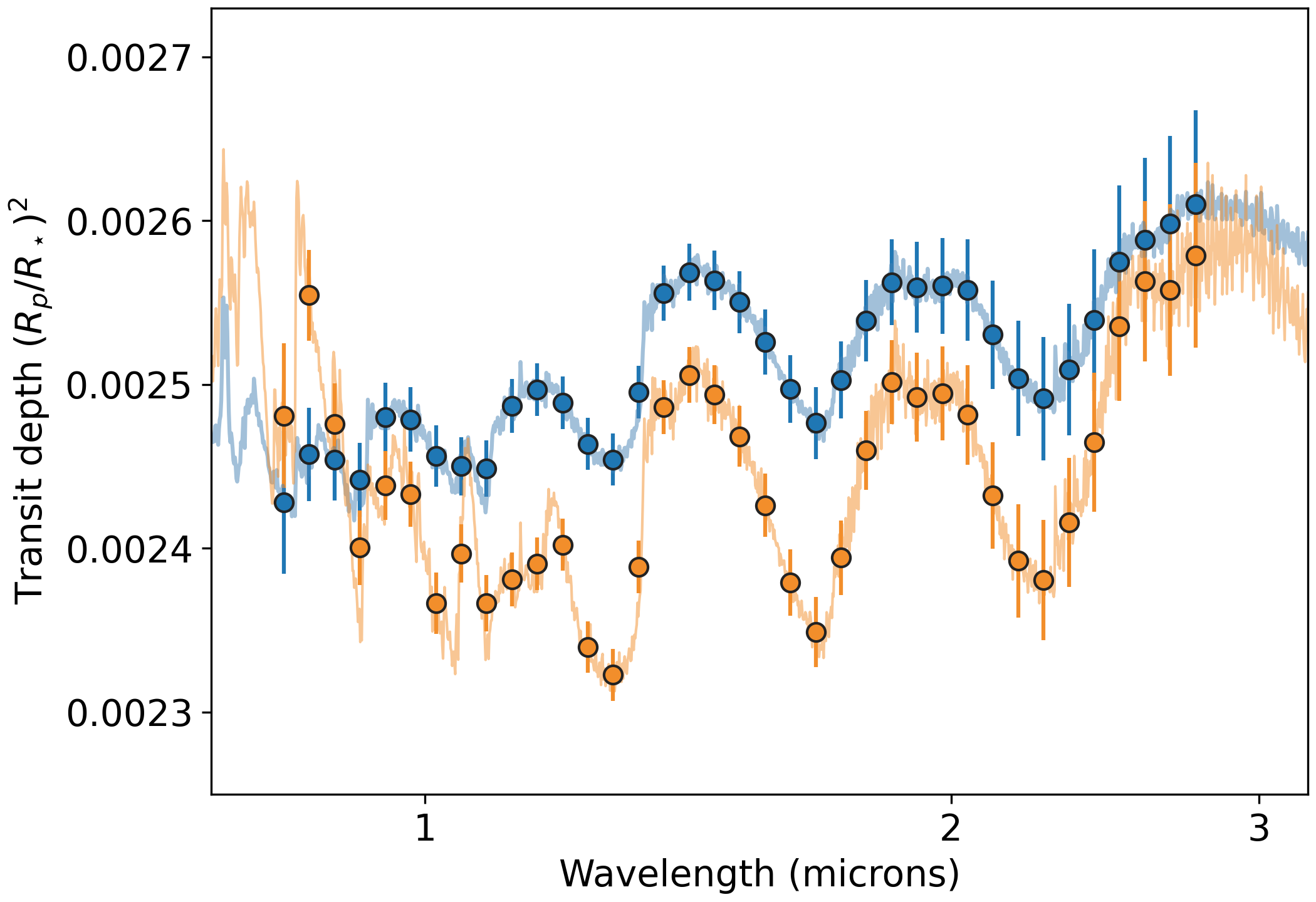}
    \caption{
    Transmission spectra for TOI-2665 b simulated with \texttt{petitRADTRANS} and respective observations using JWST NIRISS-SOSS instrument computed with \texttt{PandExo}. The orange and blue curves show [Fe/H] = 0 and [Fe/H] = 2 respectively. Both models assume atmospheric C/O = 0.55 and a cloud-free isothermal atmosphere. A single transit with NIRISS-SOSS is capable of distinguishing between solar and high metallicity atmosphere.
    }
    \label{fig:placeholder}
\end{figure}

\section{Summary}

We have reported the discovery and confirmation of TOI-2665 b, a highly irradiated sub-Saturn orbiting a metal-rich G dwarf with a period of 2.85 days. Using transit photometry from the Transiting Exoplanet Survey Satellite, ground-based follow-up observations, high-resolution imaging, and precise radial velocity measurements from the Carnegie Planet Finder Spectrograph, we measure the planet's radius and mass to be $R_p = 7.00_{-0.15}^{+0.16}~R_\oplus$ and $M_p = 57.0_{-4.2}^{+4.2} ~M_\oplus$, respectively. TOI-2665 b has a bulk density ($\rho_p \approx 0.9\,\mathrm{g\,cm^{-3}}$) implying the presence of a substantial gaseous envelope despite receiving extreme stellar irradiation ($\sim 1068~S_\oplus$). Its location within the hot Neptune desert makes it one of less than a dozen known short-period sub-Saturns to survive in this sparsely populated regime.

The super-solar metallicity of its host star suggests that TOI-2665\,b may represent the remnant of a formerly more massive giant planet that experienced significant atmospheric loss, supporting ``top down'' formation pathways proposed for hot Neptune desert inhabitants. We find no evidence for additional massive companions in the system, consistent with the predominantly single nature of desert-dwelling planets. Owing to its inflated atmosphere, high equilibrium temperature ($\sim 1600$\,K), and relatively bright host star, TOI-2665\,b is also an excellent target for atmospheric characterization with JWST. Simulated transmission spectra indicate that even a single transit observation may distinguish between solar and highly metal-enriched atmospheric compositions, offering a valuable opportunity to probe the evolutionary origins of highly irradiated giant planets.

\begin{acknowledgments}
We thank the referee for their comments which improved the quality of the paper.

Funding for the TESS mission is provided by NASA's Science Mission Directorate. We acknowledge the use of public TESS data from pipelines at the TESS Science Office and at the TESS Science Processing Operations Center. This paper includes data collected by the TESS mission that are publicly available from the Mikulski Archive for Space Telescopes (MAST). The specific observations analyzed can be accessed via \dataset[doi:10.17909/eqzn-q973]{https://doi.org/10.17909/eqzn-q973}. 

Resources supporting this work were provided by the NASA High-End Computing (HEC) Program through the NASA Advanced Supercomputing (NAS) Division at Ames Research Center for the production of the SPOC data products.

This research has made use of the Exoplanet Follow-up Observation Program (ExoFOP; DOI: 10.26134/ExoFOP5) website, which is operated by the California Institute of Technology, under contract with the National Aeronautics and Space Administration under the Exoplanet Exploration Program.

This research has made use of the NASA Exoplanet Archive, which is operated by the California Institute of Technology, under contract with the National Aeronautics and Space Administration under the Exoplanet Exploration Program.

Based on observations obtained at the Southern Astrophysical Research (SOAR) telescope, which is a joint project of the Minist\'{e}rio da Ci\^{e}ncia, Tecnologia e Inova\c{c}\~{o}es (MCTI/LNA) do Brasil, the US National Science Foundation’s NOIRLab, the University of North Carolina at Chapel Hill (UNC), and Michigan State University (MSU).

MINERVA-Australis is supported by Australian Research Council LIEF Grant LE160100001, Discovery Grants DP180100972 and DP220100365, Mount Cuba Astronomical Foundation, and institutional partners University of Southern Queensland, UNSW Sydney, MIT, Nanjing University, George Mason University, University of Louisville, University of California Riverside, University of Florida, and The University of Texas at Austin. We respectfully acknowledge the traditional custodians of all lands throughout Australia, and recognise their continued cultural and spiritual connection to the land, waterways, cosmos, and community. We pay our deepest respects to all Elders, ancestors and descendants of the Giabal, Jarowair, and Kambuwal nations, upon whose lands the MINERVA-Australis facility at Mt Kent is situated.

This research made use of \texttt{exoplanet} \citep{exoplanet:joss,exoplanet:zenodo} and its dependencies \citep{exoplanet:agol20,
exoplanet:arviz, exoplanet:astropy13, exoplanet:astropy18, exoplanet:kipping13,
exoplanet:luger18}.

MK acknowledges the support of the Natural Sciences and Engineering Research Council of Canada (NSERC), RGPIN-2024-06452. Cette recherche a été financée par le Conseil de recherches en sciences naturelles et en génie du Canada (CRSNG), RGPIN-2024-06452.
JSJ greatfully acknowledges support by FONDECYT grant 1240738 and from the ANID BASAL project FB210003.
\end{acknowledgments}

\vspace{5mm}
\facilities{Exoplanet Archive, Gaia, LCOGT, Magellan:Clay, TESS}
\software{\texttt{AstroImageJ} \citep{AstroImageJ}, \texttt{emcee} \citep{emcee}, \texttt{exoplanet} \citep{exoplanet}, \texttt{isochrones} \citep{Morton2015}, \texttt{matplotlib} \citep{Hunter2007}, \texttt{mors} \citep{Johnstone2021, Spada2013} \texttt{numpy} \citep{Harris2020}, \texttt{pandas} \citep{reback2020pandas, mckinney-proc-scipy-2010}, \texttt{PYMC} \citep{pymc2023}, \texttt{scipy} \citep{Virtanen2020}, \texttt{tess-point} \citep{tesspoint}}


\appendix
\clearpage
\counterwithin{figure}{section}  
\renewcommand{\thefigure}{A\arabic{figure}} 
\setcounter{figure}{0}

\begin{figure*}[t]
\centering
\includegraphics[width=1\linewidth]{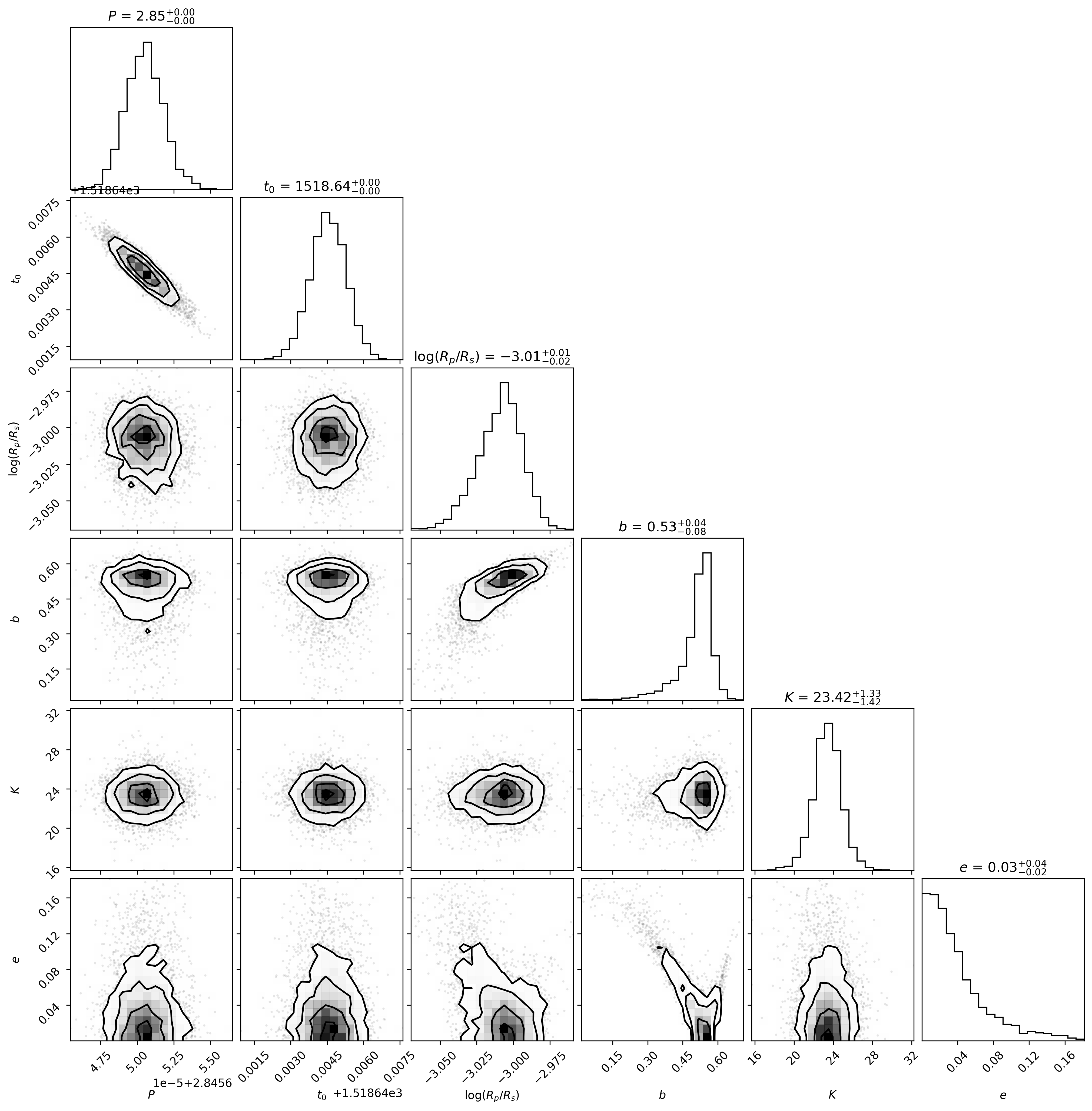}
\caption{Corner plot for key parameters ($P$, $t_{0}$, $\log{R_{p}/R_{s}}$, $b$, $K$, and $e$ based on joint transit and RV model fitting }
\label{figure: Corner plots for join fit}
\end{figure*}


\bibliography{sample701}{}
\bibliographystyle{aasjournalv7}



\end{document}